\documentclass[%
 reprint,
 amsmath,amssymb,
pra,
]{revtex4-2}

\usepackage{graphicx}
\usepackage{dcolumn}
\usepackage{bm}
\usepackage{amsmath}
\usepackage{amssymb}
\usepackage{lipsum}
\usepackage{physics}
\usepackage{bm}
\usepackage{bbold}
\usepackage{braket}
\usepackage{subcaption}
\usepackage[dvipsnames]{xcolor}
\usepackage[utf8]{inputenc}

\usepackage[colorinlistoftodos,prependcaption,textsize=scriptsize]{todonotes} %

\newcommand{\bfsfI}{\mbox{\sffamily\bfseries{I}}}
\DeclareUnicodeCharacter{03B2}{\ensuremath{\beta}}

\begin{document}

\preprint{APS/123-QED}

\title{Quantifying the breakdown of the rotating-wave approximation in single-photon superradiance}

\author{Mads Anders J{\o}rgensen$^{1}$ and  Martijn Wubs$^{1,2,3}$}
\affiliation{$^1$ Department of Photonics Engineering, Technical University of Denmark, DK-2800 Kgs. Lyngby, Denmark}%
\affiliation{$^2$ NanoPhoton - Center for Nanophotonics, Technical University of Denmark, Ørsteds Plads 345A, DK-2800 Kgs. Lyngby, Denmark}
\affiliation{$^3$ Centre for Nanostructured Graphene, Technical University of Denmark,  DK-2800 Kgs. Lyngby, Denmark}

\date{\today}

\begin{abstract}
\noindent We study quantitatively the breakdown of the rotating-wave approximation when calculating collective light emission by quantum emitters, in particular in the weak-excitation limit. Our starting point is a known multiple-scattering formalism where the full light-matter interaction leads to induced inter-emitter interactions described by the classical Green function of inhomogeneous dielectric media. When making the RWA in the light-matter interaction, however, these induced interactions differ from the classical Green function, and for free space we find a reduction of the interatomic interaction strength  by up to a factor of two. By contrast, for the corresponding scalar model the relative RWA error for the inter-emitter interaction even diverges in the near field. For two identical emitters, the errors due to the RWA in collective light emission  will show up in the emission spectrum, but not in the sub- and superradiant decay rates. In case of two non-identical emitters, also the collective emission rates will differ by making the RWA. For three or more identical emitters, the RWA errors in the interatomic interaction in general affect both the collective emission spectra and the collective decay rates. Ring configurations with discrete rotational symmetry are an interesting exception. Interestingly, the maximal errors in the collective decay rates due to making the RWA do not occur in the extreme near-field limit.   
\end{abstract}

\maketitle

\section{Introduction}

\noindent %
When resonant emitters are in close proximity, their individual single-atom emission may not describe the light emission well: collective emission can occur instead, with new resonant frequencies and decay rates corresponding to the spatial configuration of the emitters. The collective emission may be much faster and brighter than single-atom emission, which is called   superradiance~\cite{Dicke1954a,Gross:1982a}, or slower, known as subradiance~\cite{DeVoeBrewer:1996a,Guerin:2016a,Albrecht:2019a}. Such collective effects have garnered renewed interest in nanophotonics~\cite{Cong:2016a,Chang:2018a}, for 
nanolasers~\cite{Leymann:2015a,Nefedkin:2017a,Protsenko:2021a}, precision metrology~\cite{Norcia:2016a,Tang:2021a}, and in quantum information processing~\cite{Black:2005a,Hammerer:2010a,Kim:2018a,Albrecht:2019a}.

Here we focus on the interatomic interactions that enable cooperative emission, and, in particular, on the effect of the common Rotating-Wave Approximation (RWA) on these interactions. This approximation, detailed below, simplifies many calculations and is a prerequisite for several theoretical methods and models~\cite{gerry_knight_2004,Gardiner:2004a,Jacobs:2014a}. Superradiance has been studied extensively for clouds of atomic gases, often described by the Dicke model~\cite{Dicke1954a} where interatomic distances are neglected. Effects of the rotating-wave approximation have been studied for atomic clouds~\cite{FRIEDBERG:1973101,FRIEDBERG:20082514,Svidzinsky2010a}. Here instead we are mostly interested in engineered few-emitter configurations, both in free space~\cite{DeVoeBrewer:1996a} and in (in-)homogeneous dielectrics~\cite{Cong:2016a}, where interatomic interactions play a crucial role, and where the reduced complexity allows us to gain additional analytical insight (see also Ref.~\cite{Feng:2013a}).

When textbooks discuss the validity of the RWA, typically only a single emitter is considered~\cite{gerry_knight_2004}. And for a single atom (or artificial atom) coupled to a single optical mode of the electromagnetic field, phenomena beyond the RWA are known to occur when driving the atom very strongly: for one atom  driven by a single optical mode, it is known~\cite{BlochSiegert:1940a,Allen:1975} and shown in experiments~\cite{Fuchs:2009a,FornDiaz:2010} that the RWA breaks down in the ultrastrong-coupling regime, where the light-matter interaction becomes of the order of the optical transition frequency. But for weaker driving, one typically employs the celebrated Jaynes-Cummings Hamiltonian or the Quantum Maxwell-Bloch equations, that both rely on the RWA.

The validity of the RWA becomes more tricky, still for a single atom,  when off-resonant processes become important, which requires many electromagnetic modes to be considered. For example, it has been argued that the RWA should not be used for the calculation of  Lamb shifts~\cite{Agarwal:1971a, Agarwal:1973a,FRIEDBERG:1973101, MilonniKnight:1974a} and neither for thermally induced radiative shifts~\cite{Farley:1981a}. But such issues are often by-passed by assuming that these shifts are already incorporated in the observable single-atom transition frequencies~\cite{Leonardi1986a,SokhoyanAtwater:2013,Hohenester:2020a}. 

Also for the collective emission of light by several or many emitters, it is common to make the RWA in the  atom-field interaction, see for example Refs.~\cite{gerry_knight_2004,Swain_1972,Bettles:2017a,Sinha:2020a,HuLuLuZhou:2020a,Berman:2020a}. 
The individual light-emitter interactions are the same as for a single emitter, yet new issues emerge for the RWA when emitters enter each other's near fields~\cite{Swain_1972,MilonniKnight:1974a,Leonardi1986a,Carmichael:1993a,Das2018a, Hohenester:2020a}, although the (in-)accuracy of the RWA is typically not quantified.

Besides avoiding the RWA altogether~\cite{SokolovKupriyanovHavey:2011a,Li:2013a}, there is an interesting recent development to deal with these issues, where collective light emission is described using Hamiltonians with direct emitter-emitter interactions proportional to the classical Green function of the medium~\cite{AsenjoGarcia:2017a,Chang:2018a,Guimond:2019a,Fersterer:2020a}. This is a powerful formalism as it allows dielectric environments to be inhomogeneous, dispersive as well as lossy. 
It comes however at the cost of having to deal  with non-hermitian parts in the interactions since the Green function is complex-valued.

So on the one hand there is a widespread use of the hermitian RWA interaction in superradiance studies, and on the other hand an emerging formalism with non-hermitian Green-function-based light-matter interactions.  Here we are interested in the quantitative differences in the two approaches of the collective modes of few-atom systems. We will consider inhomogeneous dielectric environments, while neglecting for simplicity material dispersion and loss, and we use a formalism that is based on a canonical quantization theory~\cite{CohenTannoudji:1989a,Dalton:1996a,Dalton:1997a,Wubs:2003a}.

While in the Dicke model the emitters are assumed to be packed in a volume much smaller than the cubic wavelength, and all emission takes place in a single superradiant mode, we will also consider spatially more extended configurations, where collective emission is a multi-mode affair, both sub- and superradiant. Since it will enable us to obtain new insight, we will restrict our present study to the weak-excitation limit, also known as single-photon superradiance~\cite{Scully:2009a}, where two-level emitters can be approximated as quantum harmonic oscillators~\cite{Lehmberg:1970a,James:1993a,Scully:2009a,Svidzinsky2010a}.

We will use a quantum optical multiple-scattering formalism to describe the field dynamics. Our approach is different from most superradiance studies, since we do not integrate out the field dynamics, as in Refs.~\cite{Lehmberg:1970a,Chang:2018a,Filipovich:2021a} for example, but instead the emitter dynamics. This results in a Lippmann-Schwinger equation for the electric-field operator that can be solved analytically.   From our theory emerges a dyadic propagator  that describes interatomic interactions between different emitters in an inhomogeneous dielectric environment.  
It has been shown that when {\em not} making the RWA, this dyadic propagator is exactly identical to the classical optical Green function for any inhomogeneous environment~\cite{Wubs2004a}.

Here we make the RWA however, and we will see that a {\em different} dyadic propagator will emerge in the resulting  scattering theory, and we will study how much it differs from the classical Green function. 
We first obtain an expression for the RWA propagator for arbitrary environments, and then for free space an analytical expression for the resulting distance-dependent RWA error in the dyadic propagator. We identify at what distances the RWA error results in significant errors in predictions for collective light emission, in particular both sub- and superradiant line shifts and emission rates. Furthermore, we find that the near-field limit of the RWA error in electromagnetism  assumes a particularly simple form, especially when comparing to the very different RWA error in a corresponding scalar-wave theory. We find and explain qualitative differences in the emission by one, two, and three or more emitters.

The structure of our article is as follows: In Sec.~\ref{Sec:Formalism} we introduce our model and derive the Lippman-Schwinger equation within the RWA. In Sec.~\ref{Sec:1atom} we compute the RWA propagator both for scalar and vector fields and discuss single-emitter emission. In Sec.~\ref{Sec:Natom_scattering} we turn to collective emission by  $N$ atoms within the RWA, which we then specify in Sec.~\ref{sec:2atomCollective_emission} for two and in Sec.~\ref{Sec:three_RWA} for three and more emitters, which we argue have qualitatively different collective emission. Sec.~\ref{sec:RingSystems} presents results for ring systems which turn out to be unusually robust against making the RWA. We conclude in Sec.~\ref{Sec:summary}.

\section{Multiple-scattering formalism}\label{Sec:Formalism}
\noindent This section will closely follow the formalism of
Ref.~\cite{Wubs2004a}, but taking the RWA Hamiltonian as the starting point, rather than the full dipole interaction Hamiltonian. This will allow us to make the comparison 
of predictions of collective emission with and without making the RWA.
We will include a complete set of modes of the electromagnetic field, rather than purely the resonant modes of the atoms or that of a cavity~\cite{Bastarrachea-Magnani:2014a}, so that we can test the validity of the RWA without testing at the same time the accuracy of using only a subset of modes. 

We start by considering a general nondispersive  inhomogeneous dielectric medium characterized by the real-valued scalar dielectric function $\varepsilon({\bf r})$. The mode functions $\mathbf{f}_{\lambda}$ labelled by $\lambda$ are the solutions of the wave equation of the medium, 
\begin{subequations}\label{eq:WaveEquations}
\begin{eqnarray}\label{eq:WaveEquationVectorial}
        - \nabla \times \nabla \times \mathbf{f}_\lambda (\mathbf{r}) + \epsilon(\mathbf{r})(\omega_\lambda/c)^2\mathbf{f}_\lambda(\mathbf{r})=0,
\end{eqnarray}
with a nonnegative eigenfrequency $\omega_{\lambda}$. Since $\varepsilon({\bf r})$ is real-valued, both the real and imaginary parts of complex-valued  mode functions $\mathbf{f}_\lambda(\mathbf{r})$ are separate and degenerate solutions of Eq.~(\ref{eq:WaveEquationVectorial}). For that reason, the electromagnetic fields in the medium can be expanded in a complete set of real-valued mode functions~\cite{Wubs:2003a}, a property that we will use below.

We will also briefly consider the impact of the RWA on the scalar Green function in Sec.~\ref{sec:TheScalarField}, as it turns out to behave fundamentally different from the dyadic Green function, which is treated in Sec.~\ref{sec:TheVectorField}.
In the corresponding scalar model,  the mode functions are the solutions to the scalar wave equation 
\begin{eqnarray}\label{eq:WaveEquationScalar}
\nabla^2 f_\lambda (\mathbf{r}) + \epsilon(\mathbf{r})(\omega_\lambda/c)^2f_\lambda(\mathbf{r})=0.
\end{eqnarray}
\end{subequations}
Only the vector mode functions are written in bold font. 

\subsection{The Hamiltonian - \\Full and RWA interactions}
\noindent We consider a system of $N$ atoms or other localized quantum emitters, numbered by the index $m$, that interacts with the electromagnetic field via the dipole interaction, such that the Hamiltonian of the total system can be written as 
\begin{equation}\label{eq:TotalSymbolicHamiltonian}
    \mathcal{H} = \mathcal{H}_F + \mathcal{H}_A +\mathcal{H}_{AF},
\end{equation}
with field and atomic parts given by 
\begin{subequations}\label{eq:TotalHamiltonianExact}
\begin{eqnarray}
    \mathcal{H}_F & = & \sum_{\lambda} \hbar \omega_\lambda a^\dagger_\lambda a_\lambda, \\
    \mathcal{H}_A & = & \sum_{m=1}^N \hbar \Omega_m b^\dagger_m b_m,
\end{eqnarray}
Here we model the atoms as harmonic oscillators, which is valid in the single-photon superradiance limit~\cite{Lehmberg:1970a,James:1993a,Scully:2009a,Svidzinsky2010a}. This model is exactly solvable, apart from accurate pole approximations, see below, and this will make  the effects of the RWA most clear. However, we note in passing that our scattering formalism can also be used with two-level atoms instead~\cite{MangaRao:2007a,Yao:2009a,Ott:2013a}, where results will agree when neglecting saturation~\cite{Bouchet:2019a}; the formalism can also be generalized to  emitters strongly coupled to cavities~\cite{Hughes:2009a}, and even when the photonic environment is lossy~\cite{Delga:2014a}.  

The dipole interaction term in Eq.~(\ref{eq:TotalSymbolicHamiltonian}) is given by
\begin{equation}\label{eq:DipoleHamiltonianExactInteraction}
    \mathcal{H}_{AF} = - \sum_{m=1}^N \bm{\mu}\cdot\mathbf{F}(\mathbf{R}_m)=\sum_{m, \lambda}(b_m + b_m^\dagger)(g_{\lambda m}a_\lambda + g^*_{\lambda m}a^\dagger_\lambda),
\end{equation}
where the summation over $\lambda$ is a sum over the complete set of field modes of the electromagnetic field. We work in a (generalized) Coulomb gauge in which there is no instantaneous
interaction term between the neutral
quantum emitters, which notice each other only because they interact with
the same retarded electromagnetic field. The field operator $\mathbf{F}(\mathbf{r})$ is short-hand notation for $ {\bf D}({\bf r})/[\varepsilon_0 \varepsilon({\bf r})]$, which equals the electric-field operator everywhere except 
at the positions ${\bf R}_m$ of the emitters~\cite{Dalton:1997a,Wubs:2003a,Wubs2004a}. 

The interaction term, Eq.~(\ref{eq:DipoleHamiltonianExactInteraction}), contains four separate processes which are visualized in Fig.~\ref{fig:RWAHamiltonian}.
%
\begin{figure}[t!]
  \includegraphics[width=0.7\linewidth]{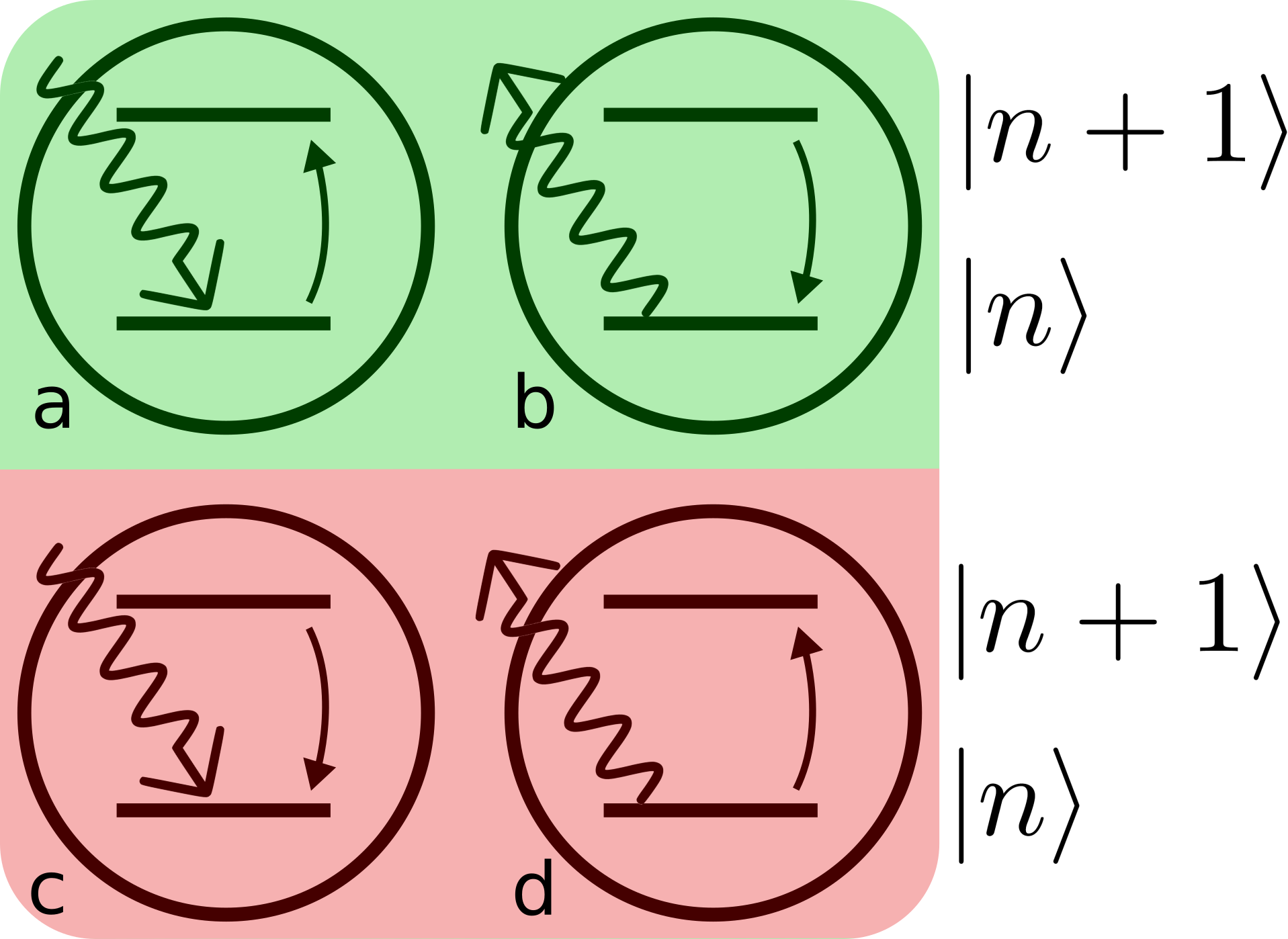}
\caption{\label{fig:RWAHamiltonian} The processes included in the full interaction, Eq.~(\ref{eq:DipoleHamiltonianExactInteraction}). (a) Excitation of an atom from state $\ket{n}$ to $\ket{n+1}$ by absorption of a photon, (b) relaxation of an atom by emission of a photon, (c) relaxation of an atom and absorption of a photon, and (c) excitation of an atom and emission of a photon. The processes on green background are called energy conserving, while those on red are strongly off-resonant and called energy non-conserving.}
\end{figure}
%
These correspond to absorption of a photon and the excitation of an atomic state, relaxation of an atomic state and the emission of a photon, relaxation of an atomic state and the absorption of a photon and, finally, excitation of an atomic state and the emission of a photon.
The first two of these processes are called resonant processes and the last pair are sometimes called energy non-conserving, counter-rotating, or ``off-shell'', and are often discarded, as the resonant terms tend to dominate in most systems.
If one indeed 
discards these two terms, what is left is the dipole interaction 
in the RWA,
\begin{equation}\label{eq:DipoleHamiltonianRWAInteraction}
    \mathcal{H}_{AF}^{\text{RWA}} = \sum_{m, \lambda} (g_{\lambda m} b_m^\dagger a_\lambda + g^*_{\lambda m} b_m a^\dagger_\lambda ).
\end{equation}
\end{subequations}
The quantized field is given by the sum of two parts,
\begin{subequations}\label{eq:PhotonicPlusandMinusFields}
\begin{equation}
      \mathbf{F}(\mathbf{r},t)=  \mathbf{F}^+(\mathbf{r},t)+  \mathbf{F}^-(\mathbf{r},t),
\end{equation}
which are related by  $\mathbf{F}^-=\left(\mathbf{F}^{+}\right)^\dagger$ and where the positive-frequency part has the mode expansion
\begin{equation}\label{eq:ElectricFieldOperatorPlusDefinition}
    \mathbf{F}^+(\mathbf{r},t)=i \sum_\lambda \sqrt{\frac{\hbar \omega_\lambda}{2 \epsilon_0}} a_\lambda(t) \mathbf{f}_\lambda(\mathbf{r}).
\end{equation}
\end{subequations}
Here the $\mathbf{f}_\lambda(\bm{r})$ are the electromagnetic modes that are the solutions to equations (\ref{eq:WaveEquationVectorial}) [or (\ref{eq:WaveEquationScalar}) for scalar waves].
The coupling constants are
\begin{equation}\label{eq:gCouplingStrengthDipoleApproximation}
    g_{\lambda m} = -i \sqrt{\frac{\hbar \omega_\lambda}{2\epsilon_0}}\bm \mu_m \cdot \mathbf{f}_\lambda(\mathbf{R}_m),
\end{equation}
where $\mathbf{R}_m$ and $\bm\mu_m$ are the spatial coordinate and fixed dipole moment of atom $m$, respectively.

\subsection{Field dynamics}
\noindent Using Eq.~(\ref{eq:DipoleHamiltonianRWAInteraction}) in the Heisenberg equation of motion,
we obtain for the atomic and field operators the coupled equations of motion
\begin{subequations}\label{Eq:equations_of_motion}
\begin{eqnarray}
    \Dot{a}_\lambda & = & - i \omega_\lambda a_\lambda - \frac{i}{\hbar} \sum_m g_{\lambda,m}^* b_m, \\
    \Dot{a}^\dagger_\lambda &= & i \omega_\lambda a_\lambda^\dagger + \frac{i}{\hbar} \sum_m g_{\lambda,m} b_m^\dagger, \\
    \Dot{b}_m & = & - i \Omega_m b_m - \frac{i}{\hbar} \sum_\lambda g_{\lambda,m} a_\lambda, \\
    \Dot{b}^\dagger_m & = & i \Omega_m b_m^\dagger + \frac{i}{\hbar} \sum_\lambda g_{\lambda,m}^* a_\lambda^\dagger.
\end{eqnarray}
\end{subequations}
We now perform a positive-time Laplace transformation, $A(\omega)=\int_0^\infty dt A(t) e^{i\omega t}$.
By adding a small positive imaginary term $i \eta$ to the 
frequency variable to ensure 
convergence of the integrals, we can 
write $\int_0^\infty \mbox{d}t\, \Dot{b}(t) e^{i(\omega + i \eta) t}= - b(t=0) - i \omega b(\omega)$, and similarly for $b^\dagger$, $a$, and $a^\dagger$.
We will let $a(t=0)$ coincide with the undisturbed field (i.e. in the absence of the atoms, equivalent to ``turning on'' the interaction at $t=0$). This undisturbed field undergoes simple harmonic time dependence, $\dot{a}^{(0)}_\lambda(t) + i \omega a^{(0)}_\lambda(t)=0$, which after Laplace transformation gives $a^{(0)}(t=0)=- i(\omega - \omega_\lambda)a^{(0)}(\omega)$.
We use this to replace $a^{(0)}(t=0)$ in the Laplace-transformed version of the system of equations~(\ref{Eq:equations_of_motion}). 
These are coupled linear algebraic equations for the frequency-dependent operators and can be solved as
\begin{subequations}
\begin{eqnarray}
    a_\lambda(\omega) & = & a_\lambda^{(0)}(\omega) + \frac{1}{\hbar(\omega - \omega_\lambda)}\sum_m^N g^*_{\lambda m} b_m(\omega), \label{eq:RWAOperatorLaplaceTransform1} \\
    a^\dagger_\lambda(\omega) & = & a_\lambda^{(0) \dagger}(\omega) - \frac{1}{\hbar(\omega + \omega_\lambda)}\sum_m^N g^*_{\lambda m} b_m^\dagger(\omega), \label{eq:RWAOperatorLaplaceTransform2} \\
    b_m(\omega) &= &\frac{ib_m(t=0)}{\omega - \Omega_m} + \frac{1}{\hbar(\omega - \Omega_m)}\sum_\lambda g_{\lambda m} a_\lambda (\omega), \label{eq:RWAOperatorLaplaceTransform3} \\
    b_m^\dagger (\omega) & = & \frac{ib_m^\dagger(t=0)}{\omega + \Omega_m} - \frac{1}{\hbar(\omega + \Omega_m)}\sum_\lambda g_{\lambda m}^* a^\dagger\lambda (\omega). \label{eq:RWAOperatorLaplaceTransform4}
\end{eqnarray}
\end{subequations}
Opposite to master-equation formalisms, where field dynamics is integrated out, we here choose to   remove nearly all explicit  dependence on the dynamics of the atomic degrees of freedom by inserting Eqs.~(\ref{eq:RWAOperatorLaplaceTransform3}) and (\ref{eq:RWAOperatorLaplaceTransform4}) into Eqs.~(\ref{eq:RWAOperatorLaplaceTransform1}) and (\ref{eq:RWAOperatorLaplaceTransform2}), respectively.
Doing so brings us to the frequency-domain expression for the photon creation and annihilation operators,
\begin{subequations}\label{eq:RWAOperatorLaplaceTransformFullForm}
\begin{eqnarray}
     a_\lambda(\omega) & = & a_\lambda^{(0)}(\omega) + \frac{i}{\hbar(\omega - \omega_\lambda)}\sum_m^N g^*_{\lambda m} \frac{b_m(t=0)}{\omega - \Omega_m} \nonumber \\ & + 
     &\frac{1}{\hbar^2 (\omega - \omega_\lambda)}\sum_{\lambda' m}g^*_{\lambda m} \frac{1}{\omega - \Omega_m}g_{\lambda' m}a_{\lambda'}(\omega)
\end{eqnarray}
and
\begin{eqnarray}
     a_\lambda ^\dagger (\omega) & = & a_\lambda^{(0) \dagger}(\omega) - \frac{i}{\hbar(\omega + \omega_\lambda)}\sum_m^N g_{\lambda m} \frac{b_m^\dagger (t=0)}{\omega +\Omega_m} \nonumber\\ & + 
     &\frac{1}{\hbar^2 (\omega + \omega_\lambda)}\sum_{\lambda' m}g_{\lambda m} \frac{1}{\omega + \Omega_m}g_{\lambda' m}^* a_{\lambda'}^\dagger (\omega).
\end{eqnarray}
\end{subequations}
We note that these expressions, obtained within the RWA,  are quite a bit simpler than Eqs.~(10a-b) in Ref.~\cite{Wubs2004a}, which were found with the full interaction of Eq.~(\ref{eq:DipoleHamiltonianExactInteraction}).

Inserting Eqs.~(\ref{eq:RWAOperatorLaplaceTransformFullForm}) into Eqs.~(\ref{eq:PhotonicPlusandMinusFields}) enables us to find a self-consistent Lippmann-Schwinger equation for the field 
\begin{eqnarray}\label{eq:RWAFieldOperatorPlusMinusFreeSourceScattering}
\mathbf{F}^\pm(\mathbf{r},&& \omega) = \mathbf{E}^{\pm(0)}(\mathbf{r}, \omega) + \sum_m \mathbf{K}^\pm(\mathbf{r}, \mathbf{R}_m, \omega)\cdot \mathbf{S}^\pm_m(\omega)\nonumber \\
 & & + \sum_m \mathbf{K}^\pm(\mathbf{r}, \mathbf{R}_m, \omega)\cdot \mathbf{V}_m^\pm (\omega) \cdot \mathbf{F}^\pm (\mathbf{R}_m, \omega).
\end{eqnarray}
This Eq.~(\ref{eq:RWAFieldOperatorPlusMinusFreeSourceScattering}) was obtained with the RWA interaction and  describes the positive- and negative-frequency contributions to the electric-field operator separately [because the superscripts $\pm$ do not mix in Eq.~(\ref{eq:RWAFieldOperatorPlusMinusFreeSourceScattering})].
By contrast, using the full interaction returns a single self-consistent equation where the full field enters, including both positive- and negative-frequency contributions (see Ref.~\cite{Wubs2004a}). Eq.~(\ref{eq:RWAFieldOperatorPlusMinusFreeSourceScattering}) is a self-consistent equation for the  unknown fields $\mathbf{F}^\pm(\mathbf{r}, \omega)$, as they appear both on the left- and on the right-hand side of the equation. The field can thus be solved iteratively, in ever higher orders of the scattering potential, which explains the name ``multiple-scattering formalism'' of this section. Since we work in the RWA and consider positive frequencies $\omega$, we will focus on the rotating part $\mathbf{F}^+(\mathbf{r}, \omega)$ of the total field of Eq.~(\ref{eq:RWAFieldOperatorPlusMinusFreeSourceScattering});  the anti-rotating part can be obtained from it from the identity $\mathbf{F}^{-}(\mathbf{r}, \omega) = \left[\mathbf{F}^+(\mathbf{r}, - \omega^{*})\right]^{\dag}$.

Each term in Eq.~(\ref{eq:RWAFieldOperatorPlusMinusFreeSourceScattering}) corresponds to a different contribution to the field at position $\mathbf{r}$.
The first term on the right-hand side
is the undisturbed field which has not been altered by the presence of the atoms, and corresponds to Eq.~(\ref{eq:ElectricFieldOperatorPlusDefinition}), and its hermitian conjugate, with $a_\lambda\rightarrow a_\lambda^{(0)}$.
The second term is interpreted as light originating from a source term at one of the atomic positions and subsequently propagating to $\mathbf{r}$.
The last term is a contribution from the whole field scattered on the $m$'th atom, before propagating to $\mathbf{r}$.

In our RWA formalism,  the atomic potentials $\mathbf{V}_m^\pm$ and source terms $ \mathbf{S}_m^{\pm}$ are given by 
\begin{equation}
    \mathbf{V}_m^\pm (\omega) = 
 \hat{\bm \mu}_m V_m^\pm(\omega)\hat{\bm \mu}_m =     
    \hat{\bm \mu}_m \left(\frac{\mu_m^2\omega^2}{ \hbar \epsilon_0 c^2}\right)\frac{1}{\omega \mp \Omega_m} \hat{\bm \mu}_m,
\end{equation}
and 
\begin{subequations} \label{eq:RWADyadAndVectorDefinitions}
\begin{equation}
    \mathbf{S}_m^+ (\omega)=\hat{\bm \mu}_m S^+_m(\omega) = \hat{\bm \mu}_m \left( \frac{- i \mu_m \omega^2}{ \epsilon_0 c^2} \right) \frac{b_m(0)}{\omega - \Omega_m},
\end{equation}
\begin{equation}
    \mathbf{S}_m^- (\omega)=\hat{\bm \mu}_m S^-_m(\omega) = \hat{\bm \mu}_m \left( \frac{ i \mu_m \omega^2}{ \epsilon_0 c^2} \right) \frac{b_m^\dagger(0)}{\omega + \Omega_m},
\end{equation}
\end{subequations}
 respectively.  The final unknowns in Eq.~(\ref{eq:RWAFieldOperatorPlusMinusFreeSourceScattering}) are the dyadics $\mathbf{K}^{\pm}$, interpreted above as propagators of the EM field, which are given by
\begin{equation}\label{eq:KPlusMinusRWADefinition}
      \mathbf{K}^\pm(\mathbf{r}, \mathbf{r}', \omega) = \frac{c^2}{2}\sum_{\lambda}\frac{\mathbf{f}_\lambda(\mathbf{r})\mathbf{f}_\lambda^*(\mathbf{r'})}{\omega \mp \omega_\lambda}\frac{\omega_\lambda}{\omega^2}.
\end{equation}
So given a complete set of optical modes of an arbitrary inhomogeneous dielectric environment and their eigenfrequencies, these two RWA propagators can be calculated. 

For comparison,   the Lippmann-Schwinger equation obtained in Ref.~\cite{Wubs2004a} when using the full interaction features the propagator 
\begin{equation}\label{eq:KDefinition}
      \mathbf{K}(\mathbf{r}, \mathbf{r}', \omega) = c^2\sum_{\lambda}\frac{\mathbf{f}_\lambda(\mathbf{r})\mathbf{f}_\lambda^*(\mathbf{r'})}{\omega^2 - \omega_{\lambda}^{^2}}\frac{\omega_{\lambda}^{2}}{\omega^2},
\end{equation}
and it could be shown that (everywhere except for $\mathbf{r}=  \mathbf{r}'$ ~\cite{Wubs2004a}) this propagator $\mathbf{K}$ is identical to the classical electromagnetic dyadic Green function $\mathbf{G}$ of the inhomogeneous medium, defined by the equation 
\begin{equation}
- \nabla \times \nabla \times \mathbf{G}(\mathbf{r}, \mathbf{r}', \omega) + \epsilon(\mathbf{r})(\omega/c)^2\mathbf{G}(\mathbf{r}, \mathbf{r}', \omega)= \bfsfI \delta^{3}(\mathbf{r}- \mathbf{r}').
\end{equation} 
Using this identity and Eqs.~(\ref{eq:KPlusMinusRWADefinition}) and (\ref{eq:KDefinition}), we now find the  identity for our RWA propagators
\begin{equation}\label{eq:GisKminusK}
\mathbf{G}(\mathbf{r}, \mathbf{r}', \omega) = \mathbf{K}^{+}(\mathbf{r}, \mathbf{r}', \omega) - \mathbf{K}^{-}(\mathbf{r}, \mathbf{r}', \omega), \quad \mbox{for}\,\, \mathbf{r} \ne  \mathbf{r}'.    
\end{equation}
Here the two RWA propagators are the less familiar objects, but their difference gives the classical electromagnetic Green function that is known explicitly for various model dielectric environments~\cite{Tai:1994a,Sondergaard2019a}. However, given a real-space representation of a Green function $\mathbf{G}$ in which the sum over modes has been performed explicitly, there seems to be no obvious way to split it into its rotating part $\mathbf{K}^{+}$ and its anti-rotating part $\mathbf{K}^{-}$. Identifying these two components thus involves some real work even for explicitly known Green functions $\mathbf{G}$. This can be used as an argument against making the RWA; another such argument would be any non-negligible inaccuracies due to making the RWA, a point that we elucidate below.      
 
An important difference between $\mathbf{K}^{+}(\omega)$ and $\mathbf{K}^{-}(\omega)$ is that for the positive frequencies $\omega$ that we consider, the $\mathbf{K}^{+}(\omega)$ is complex-valued while $\mathbf{K}^{-}(\omega)$ is real-valued, as follows from the following argument: recall that the frequency $\omega$ in the denominator of Eq.~(\ref{eq:KPlusMinusRWADefinition}) is understood to have an infinitesimally small positive imaginary part $i \eta$. We could have chosen a basis of real-valued mode functions to expand these propagators in (recall the discussion below Eq.~(\ref{eq:WaveEquationVectorial})). Therefore, the imaginary parts of the propagators will originate solely from the delta-function in the Dirac identity $\lim_{\eta \to 0} (\omega \mp \omega_{\lambda} + i \eta)^{-1} = \PV (\omega \mp \omega_{\lambda})^{-1} - i \pi \delta(\omega \mp \omega_{\lambda})$. It follows that indeed  $\mathbf{K}^{+}(\omega)$ is complex-valued while $\mathbf{K}^{-}(\omega)$ is real-valued for arbitrary inhomogeneous nondispersive and lossless media. Physical consequences will be discussed below.

 The Lippmann-Schwinger equation~(\ref{eq:RWAFieldOperatorPlusMinusFreeSourceScattering}) is a main result of this paper which can be used to study single-atom  as well as collective light emission in arbitrary inhomogeneous dielectrics, within the RWA.  We will use Eq.~(\ref{eq:GisKminusK})  for quantitative comparisons with results obtained with the full light-matter interaction. 

\section{Single-atom emission characteristics}\label{Sec:1atom}
\noindent We will now first examine the influence of the RWA on the  emission characteristics of a single atom  interacting with the electromagnetic field at a position $\mathbf{R}$ in an inhomogeneous photonic medium. 

We can iteratively insert the full field on the left-hand side of equation (\ref{eq:RWAFieldOperatorPlusMinusFreeSourceScattering}) into the scattered term on the right-hand side, and collect all terms that arise due to a scattering of the free field, as well as the terms arising due to the source term.
The scattered field is described by 
\begin{equation}
     \mathbf{F}_{\text{scatter}}^{+}(\mathbf{r}) =\mathbf{K}^{+}(\mathbf{r}, \mathbf{R})\cdot \mathbf{T}^{+}(\omega)\cdot \mathbf{E}^{(0)+}(\mathbf{R}),
\end{equation}
where the T-matrix or scattering matrix $\mathbf{T}^{+}$, which takes into account all orders of scattering events, has the dyadic form $\mathbf{T}^{+}(\omega)=  \hat{\bm \mu}T^{+}(\omega)\hat{\bm \mu}$, with the scalar  
\begin{equation}\label{eq:SingleAtomScatteringMatrix}
 T^{+}(\omega)=   \frac{V^{+}(\omega)}{1- \hat{\bm \mu}\cdot \mathbf{K}^{+}(\mathbf{R}, \mathbf{R},\omega)\cdot \hat{\bm \mu}\, V^{+}(\omega) }.
\end{equation}
The source term is given by 
\begin{equation}\label{eq:SingleAtomFSourceTerm}
      \mathbf{F}_{\text{source}}^{+}(\mathbf{r},\omega) = \frac{ \mathbf{K}^{+}(\mathbf{r}, \mathbf{R},\omega)\cdot \mathbf{S^{+}(\omega)}}{1- \hat{\bm \mu}\cdot \mathbf{K}^{+}(\mathbf{R}, \mathbf{R}, \omega)\cdot \hat{\bm \mu}\, V^{+}(\omega) },
\end{equation}
and it describes emission of light 
by the atom at ${\bf R}$, followed by field propagation to $\mathbf{r}$. The denominators of the scattering and source terms are identical, so both terms have the same resonances.

\subsection{Single-atom spontaneous-emission rate and the RWA}\label{sec:TheAtomicSourceTerm}
%
\noindent We now analyze the properties of the source term Eq.~(\ref{eq:SingleAtomFSourceTerm}). To begin with, from its poles (i.e. the zeroes of the denominator in the complex-frequency plane) we can learn the resonance condition for single-atom light emission.
By multiplying both the numerator and the denominator by a factor of $(\omega - \Omega)$, we find that the resonance frequency is found by the solution to $\omega-\Omega+X^{+}(\omega)=0$,
 where 
\begin{equation}\label{eq:RWAPoleShift}
    X^{+}(\omega)=\hat{\mu}\cdot \mathbf{K}^{+}(\mathbf{R},\mathbf{R},\omega)\cdot\hat{\mu} \frac{\mu^2 \omega^2}{ \hbar \epsilon_0 c^2}.
\end{equation}
If we assume that the resonance condition is close to the bare atomic transition frequency, then we can find the approximate resonance frequency to be $\Omega_{1}^{+}=\Omega+X^{+}(\Omega)$ (by the pole approximation). Unlike $\Omega$, the resonance frequency $\Omega_{1}^{+}$ is complex-valued, dressed by the light-matter interaction.
The real part of $X^{+}(\Omega)$ contributes a shift to the resonance frequency of the system, while the imaginary part corresponds to the  spontaneous-decay rate $\Gamma^{+}({\bm r}, \omega)$ which depends on position, dipole orientation, and frequency. We find $\Gamma^{+}({\bm r}, \omega)=-2\Im\left[X^{+}(\Omega)\right]=\pi/(\hbar \epsilon_0)\sum_\lambda \vert\bm \mu\cdot\mathbf{f}_\lambda(\bm r)\vert^2 \omega_\lambda \delta(\Omega - \omega_\lambda)$.

This RWA prediction of the single-atom spontaneous-decay rate $\Gamma^{+}({\bm r}, \omega)$ coincides exactly with the analogous expression $\Gamma({\bm r}, \omega)$ obtained by making the pole approximation in the exact formalism~\cite{Wubs2004a}. So this justifies making the RWA if one is solely interested in single-atom decay rates. 
For the special case of emission into free space, Milonni and Knight~\cite{MilonniKnight:1974a} and Leonardi et al.~\cite{Leonardi1986a} also found the same single-atom spontaneous-decay  rates with and without making the RWA,  using different methods. Leonardi et al.~\cite{Leonardi1986a} interpreted that the full and RWA interactions should indeed give the same spontaneous-emission rates because it is an on-shell (i.e. energy-conserving)  process. Below we will see whether such an interpretation also holds for  multi-atom collective emission rates.

\subsection{RWA free-space propagator in scalar model}\label{sec:TheScalarField}
\noindent As the second property of the source term  Eq.~(\ref{eq:SingleAtomFSourceTerm}), we  evaluate the propagator $\mathbf{K^{^+}(\mathbf{r},\mathbf{R},\omega)}$. 
The propagator depends on the photonic environment considered. Since collective emission is often studied in free space and already for free space we will find differences between the RWA propagator and the exact propagator (i.e. based on the full interaction), it is most instructive to focus on a free-space environment in the following.
 
For free space, the modes are given by a continuum of  transverse plane waves with linear dispersion relation where $\omega_{\bf k}=k c$ and we can specify a relevant class of summation over modes as
\begin{eqnarray}\label{eq:SumOfModesRewriting}
   && \sum_\lambda \mathbf{f}_\lambda(\mathbf{r})\mathbf{f}^*_\lambda(\mathbf{r'}) p(\omega_{\lambda})=\sum_{{\bf k},s}\frac{p(\omega_{{\bf k}})}{V}e^{\mathbf{k}\cdot(\mathbf{r}-\mathbf{r'})}\hat{\bf e}_{\mathbf{k},s}\hat{\bf e}_{\mathbf{k},s}^* \nonumber\\
   &&=\frac{1}{(2 \pi)^3}\int \text{d}^3\mathbf{k}\, p(\omega_{{\bf k}}) \left[\bfsfI - \hat{\bf k}\otimes\hat{\bf k}\right]e^{i \mathbf{k}\cdot (\mathbf{r}-\mathbf{r'})},
\end{eqnarray}
where $V$ is a quantization volume, $p(\omega_{{\bf k}})$ an arbitrary function that we specify below, and $\bfsfI$ the unit matrix in 3D. In the last equality we summed over the two independent polarization directions and the $\hat{\bf k}\otimes\hat{\bf k}$ term accounts for the transverse nature of the free-space optical modes.

Before studying the free-space electromagnetic propagator in the RWA, it is instructive to first consider the effect of the RWA on the corresponding scalar model, recall Eq.~(\ref{eq:WaveEquationScalar}), with scalar plane waves as the complete set of modes. 
The scalar field is not sufficient to describe collective effects in sub-wavelength systems, but it is sometimes used in superradiance studies~\cite{Scully:2009a} and useful for us to study effects of the counter-rotating terms for few-emitter systems, complementing investigations for atomic clouds~\cite{FRIEDBERG:20082514}.  If only the scalar field is considered, we leave out the term proportional to $\hat{\bf k}\otimes\hat{\bf k}$ in Eq.~(\ref{eq:SumOfModesRewriting}).
In order to directly evaluate $K^{+}(\mathbf{r},\mathbf{r}',\omega)$ for the scalar field, we can combine Eq.~(\ref{eq:KPlusMinusRWADefinition}) and the part of Eq.~(\ref{eq:SumOfModesRewriting}) proportional to~$\bfsfI$ to get
\begin{eqnarray}\label{eq:RWAScalarIntegralEquation1}
K^{+} &&(\mathbf{r},\mathbf{r}',\omega)=\frac{c^2}{2 \cdot(2 \pi)^3}\int \text{d}^3\mathbf{k}\frac{k c}{\omega^2}\frac{e^{i\mathbf{k}\cdot(\mathbf{r}-\mathbf{r}')}}{\omega-k c}\nonumber\\
=&& -\frac{1}{2\cdot(2 \pi)^2 i k_0^2 R}\int_0^\infty \text{d}k \frac{k^2}{k-k_0}\left(e^{ikR}-e^{-ikR}\right).
\end{eqnarray}
Here the wave vector magnitude $k$ is the remaining integration variable resulting from summing over all modes and integrating over all directions. Furthermore, we introduced  $k_0=\omega/c$ and  $R=\vert\mathbf{r}-\mathbf{r}'\vert$.
If we had not made the RWA, then at this point we could rewrite the integral to become of the form $\int_{-\infty}^\infty \text{d}k$, which would greatly simplify things~\cite{Wubs2004a}. So although we found that intermediate expressions such as Eq.~(\ref{eq:RWAOperatorLaplaceTransformFullForm}) were simpler by making the RWA, here we find instead that  the RWA makes the math more complicated in our formalism.

\begin{figure*}[t!]
    \centering
    \begin{subfigure}[b]{0.2\textwidth}
        \centering
    \includegraphics[height=1.2in]{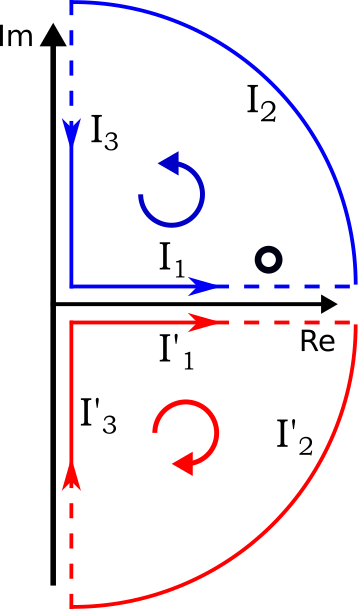}
        \caption{}\label{fig:ComplexCoordinateSystemFirstFourthQuadrant}
    \end{subfigure}%
    \begin{subfigure}[b]{0.6\textwidth}
        \centering
        \includegraphics[height=1.2in]{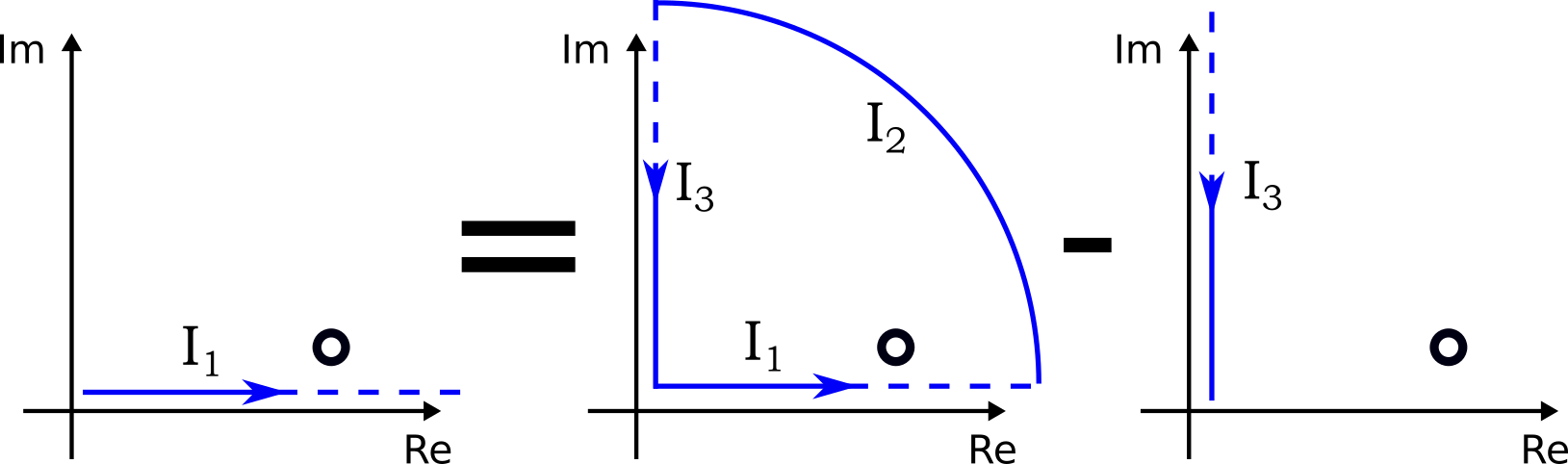}
        \caption{} \label{fig:ComplexIntegralReqriting2}
    \end{subfigure}
    \caption{(a) Schematic visualization of the complex contour integration with a pole in the first quadrant (marked by the black ring), where we 
    integrate along the positive real axis by enclosing a contour around the first (blue) or fourth (red) quadrants. (b) Visualization of the rewriting used to evaluate the integral along the positive real axis in the first quadrant. Since the integral along the arc shown in solid vanishes when the arc radius goes to infinity, the integral must be equal to the contour around the pole minus the integration down the imaginary axis. The dashed lines indicate that these integrations stretch to or from infinity along the respective axes.}\label{fig:ComplexIntegralRewriting}
\end{figure*}

\begin{figure}[h!]  
\includegraphics[width=\linewidth]{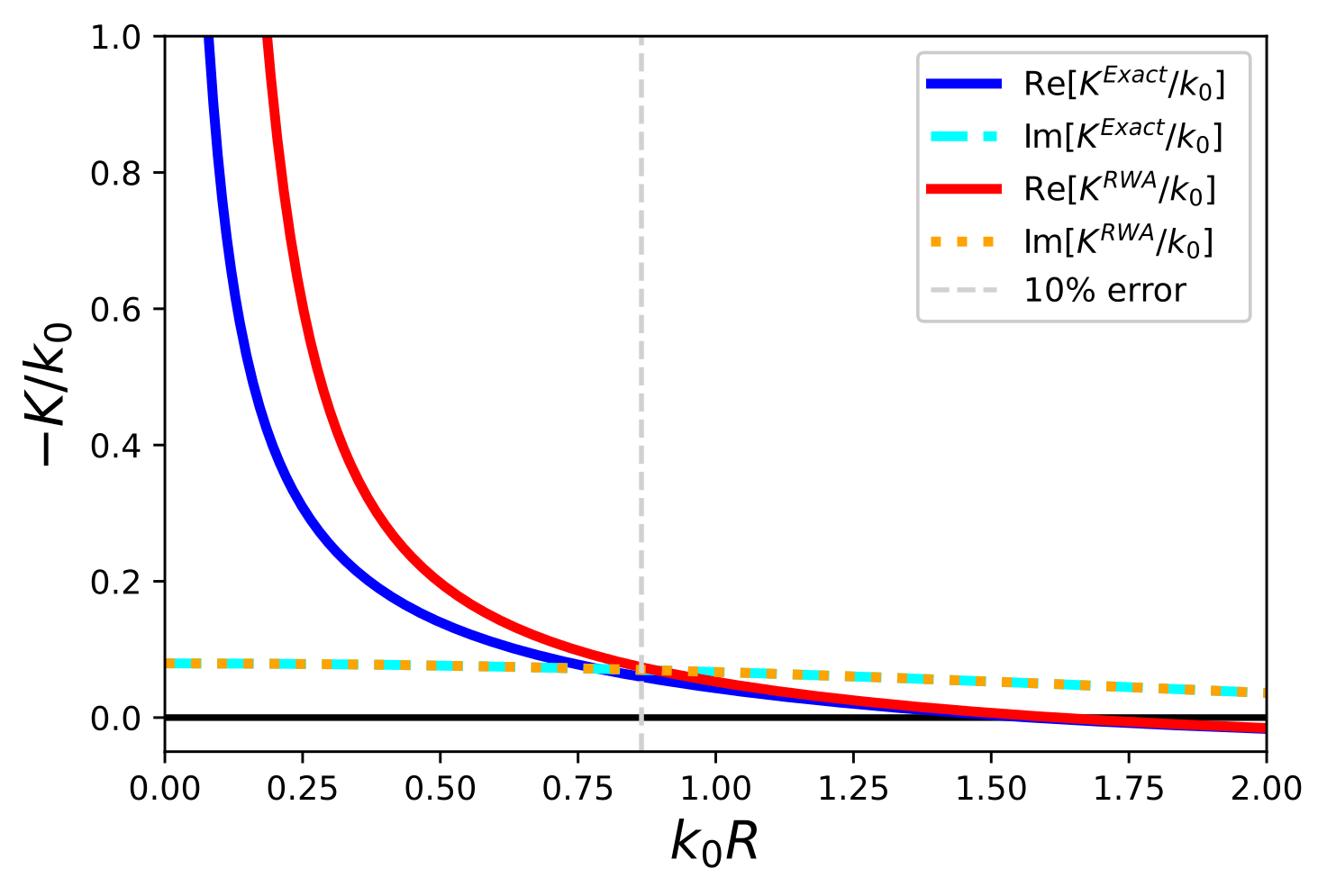}
  \caption{Real and imaginary parts of the normalized scalar Green's function,
  $K(\mathbf{R},k_0)/k_0$, evaluated with the full  and with the RWA interaction, see Eq.~(\ref{eq:RWAScalarFieldKEquation2}). The vertical dashed line at  $k_0R\approx0.87$ indicates where the real part of the relative error has reached 10\text{\%}. 
  }
  \label{fig:XRWAComparisonRealImaginaryCutoff}
\end{figure}

To proceed within the RWA, we will carry out the $k$-integration by extending the integration to the complex plane by letting $k_0\rightarrow k_0+i\epsilon$ and using complex contour integration, while taking into account the fact that we are not integrating across the entire real axis.
The differing signs in the exponential phase functions mean that we need to close contours in the first and fourth quadrants, respectively, in order to ensure that the integrals along the arc of the contours will vanish. This is illustrated in Fig.~\ref{fig:ComplexCoordinateSystemFirstFourthQuadrant}.
%
Since we are integrating in a single quadrant per term, as opposed to a full half-plane when not making the RWA, we need to subtract the integral along the imaginary axis. Symbolically, this can be expressed as $\oint_Q=I_1+I_2+I_3 \Longleftrightarrow I_1=\oint_Q-I_2-I_3=\oint_Q-I_3$, as $I_2$ vanishes due to the choice of contour. 
This rewriting is illustrated in Fig.~\ref{fig:ComplexIntegralReqriting2}.
The pole is located in the first quadrant, and so integration of the term proportional to $e^{ikR}$ in Eq.~(\ref{eq:RWAScalarIntegralEquation1}) will have a contribution from the contour integral, while the integration in the fourth quadrant will consist only of the integral along the imaginary axis, $I_1'=-I_3'$. Doing so gives us
\begin{eqnarray}\label{eq:RWAScalarFieldKEquation1}
    K^{+}(R,\omega)&= & -\frac{e^{i k_0 R}}{4 \pi R}\nonumber\\
    & - & \frac{1}{(2 \pi)^2 k_0 R} \int_0^\infty \mbox{d}k\, k^2 \frac{e^{-kR}}{k^2 + k_0^2},
\end{eqnarray}
with $k_0 = \omega/c$. We have hereby determined the scalar version of the free-space propagator that in the RWA describes light propagation by an emitter in Eq.~(\ref{eq:SingleAtomFSourceTerm}). We can immediately identify the first term on the right-hand side of Eq.~(\ref{eq:RWAScalarFieldKEquation1}), being the residue of the pole, as the scalar free-space Green function for the Helmholtz equation~\cite{NovotnyHecht:2000a}, $G_0(R,\omega)=-e^{i k_0 R}/4 \pi R$. Therefore we can identify the second term, produced by the integrations along the imaginary axis, as the error term introduced by making the RWA. We have checked that this error is indeed given by $K^{-}(R,\omega)$, in agreement with Eq.~(\ref{eq:GisKminusK}). This error varies with distance and with frequency, and is indeed purely real-valued, as expected from the discussion below Eq.~(\ref{eq:GisKminusK}). 
 The integral in Eq.~(\ref{eq:RWAScalarFieldKEquation1}) can not be simplified further and must be evaluated numerically. To evaluate and visualize the error, we note that from  
Eq.~(\ref{eq:RWAScalarFieldKEquation1}) we can write 
\begin{equation}\label{eq:RWAScalarFieldKEquation2}
       K^{+}(R,\omega)/k_0  =  -\frac{ e^{i s}}{4 \pi s} 
    -  \frac{1}{(2 \pi s)^2 } \int_0^\infty \mbox{d}u  \frac{u^2 e^{-u}}{u^2 + s^2},
\end{equation}
i.e. a rescaling of the RWA propagator with the right-hand side expressed completely in terms of the dimensionless variable $s = k_0R$.
In Fig.~\ref{fig:XRWAComparisonRealImaginaryCutoff} 
we show the real and imaginary parts of the exact and the RWA scalar propagators, as a function of $k_0R$. As the figure illustrates, 
the exact and RWA expressions for the imaginary parts of the propagators agree completely. By contrast, the real parts of the propagators both  diverge. It is not surprising that they diverge, but it is surprising how: the RWA scalar propagator diverges more strongly than the exact scalar one in the limit $k_0 R \rightarrow 0$. 
The relative error grows as the distance is reduced.  Only for distances $k_0R > 0.87$ is the error due to making the RWA  smaller than ten percent,  i.e. $\Re\left[(K_{\rm RWA} - K_{\rm Exact})/K_{\rm Exact}\right]<0.1$.  
The discrepancy is purely a consequence of neglecting all  non-rotating interactions with the free-space optical modes - there is no external driving field.

Now that we evaluated the scalar free-space propagator Eq.~(\ref{eq:RWAScalarFieldKEquation2}), we can also take the $R \rightarrow 0$ limit  with Eq.~(\ref{eq:RWAPoleShift}) to determine the radiative shift and the spontaneous-decay rate due to the atomic interaction with the electromagnetic field. We find $\Omega_{\rm scalar}^{\rm RWA} =\Omega+X_{\rm scalar}^{\rm RWA}(\Omega)$ (again in the pole approximation),
so that the complex radiative shift of the single-atom frequency is given by
 $X^{\text{RWA}}_{\text{scalar}}=X^{\text{Exact}}_{\text{scalar}}+X^{\text{Error}}_{\text{scalar}}$, 
where
$    X^{\text{Exact}}_{\text{scalar}}\underset{s\to 0}{=}-\frac{\mu^2k_0^3}{\hbar \epsilon_0}\frac{e^{i s}}{4 \pi s} $. Since $ X^{\text{Exact}}_{\text{scalar}}$ scales as $(k_0R)^{-1}$, the radiative shift actually diverges and these divergencies are well known.  
The RWA error gives rise to an additional divergent shift
\begin{equation}\label{eq:ScalarRWAXEquations3}
    X^{\text{Error}}_{\text{scalar}} \underset{s\to 0}= 
    - \frac{\mu^2 k_0^3}{\hbar \epsilon_0}\frac{1}{(2\pi s)^2 }
    \int_{0}^{\infty}\mbox{d}u\,   \frac{u^2 e^{-u}}{u^2 + s^2}.
\end{equation}
Here the dimensionless integral converges to unity in the limit $s=k_0 R\rightarrow0$.
Since the prefactor scales as $(k_0R)^{-2}$, whereas $ X^{\text{Exact}}_{\text{scalar}}$ scales as $(k_0R)^{-1}$, the error term diverges faster than the exact radiative shift, $\Re[X^{\text{Error}}_{\text{scalar}} ]$. 

Here we will not discuss how these infinities can be regularized, leading to the single-atom Lamb shift, and refer to literature~\cite{Bethe:1947a,FRIEDBERG:20082514,Farley:1981a, ScheelBuhmann2008a}. %
Instead we adopt the common pragmatic approach, as used in Refs.~\cite{Leonardi1986a,SokhoyanAtwater:2013,Hohenester:2020a} for example:  for the single-atom shifts we assume that a more microscopic approach (which we will not pursue here) can regularize the infinite shifts such that $\Omega + \Re[X(\Omega)]$ corresponds to the experimentally observable real parts of the single-atom resonance frequencies. The regularization will  be different with and without the RWA, but the outcome of the pragmatic approach in both cases is the replacement of a divergent frequency by the same observable resonance frequency. So as far as single-emitter resonance frequencies are concerned, these are identical by construction in our approach.

For the single-atom spontaneous-emission rates with and without the RWA, the situation is different than for the energy shifts: since the error Eq.~(\ref{eq:ScalarRWAXEquations3}) is purely real-valued, there is no change to the scalar free-space spontaneous-decay rate of the single atom when making the RWA, as a special case of what we found for arbitrary nondispersive photonic media in Sec.~\ref{sec:TheAtomicSourceTerm}. So these single-atom spontaneous-emission rates with and without the RWA are identical, but not identical by construction.

\subsection{RWA electromagnetic propagator }\label{sec:TheVectorField}
\noindent After analyzing the RWA popagator in a scalar model, let us now analogously determine the RWA version of the dyadic free-space electromagnetic propagator, to see how much it differs from the known dyadic Green function that is found when using the full interaction. We  take a closer look at the entire vectorial field in Eq.~(\ref{eq:SumOfModesRewriting}).
We can rewrite the angular part of the integral via~\cite{deVries1998a}
\begin{equation}\label{eq:DeVriesAngularIntRewriting}
\int d\hat{\bf k}\, e^{i \mathbf{k}\cdot \mathbf{r}} \hat{\bf k}\otimes\hat{\bf k} = -\frac{2 \pi}{k^2}  \nabla \otimes \nabla \left(\frac{e^{ikr}-e^{-ikr}}{i k r}\right).
\end{equation}
Combining this with Eqs.~(\ref{eq:KPlusMinusRWADefinition}), (\ref{eq:SumOfModesRewriting}) and~(\ref{eq:RWAScalarIntegralEquation1}) gives 
\begin{eqnarray}
&&\mathbf{K}^{+}(R,k_0)=-\frac{1}{2\cdot(2\pi)^2k_0^2}\nonumber\\
 \times&&\int_0^\infty \text{d}k\left[\bfsfI + \frac{1}{k^2}\nabla\otimes\nabla\right] \frac{k^2}{k-k_0}\frac{e^{ikR}-e^{-ikR}}{iR}.
\end{eqnarray}
Just like in the scalar case, we can rewrite the integration as a contour integral minus an integral along the imaginary axis, recall Fig.~\ref{fig:ComplexIntegralRewriting}.
This again requires closing contours in the first and fourth quadrants for the positive- and negative-phase exponential functions, respectively.
Doing so gives 
\begin{eqnarray}\label{Eq:full_plus_error1}
\mathbf{K}^{+} &&(R,k_0)=-\left[\bfsfI + \frac{1}{k_0^2}\nabla\otimes\nabla\right]\frac{e^{ik_0R}}{4 \pi R} \\
&& -\frac{1}{(2\pi)^2k_0}\int_0^\infty \left[\bfsfI - \frac{1}{k^2}\nabla\otimes\nabla\right]\frac{e^{-kR}}{R}\frac{k^2}{k^2+k_0^2}, \nonumber
\end{eqnarray}
where for $R\neq 0$ we identify the first term as the free-space dyadic Green function, $\mathbf{G}_0(R,k_0)=-\left[\bfsfI + \frac{1}{k_0^2}\nabla\otimes\nabla\right]e^{ik_0R}/(4 \pi R) $, 
which after working out the derivatives has the familiar form~\cite{NovotnyHecht:2000a}
\begin{eqnarray}\label{eq:DyadicGreenFunctionFreeSpace}
\mathbf{G}_0(R,k_0)/k_0 & = & -\frac{e^{i s}}{4 \pi s}\left[  \left( \bfsfI - \hat{\bf r}\otimes \hat{\bf r}\right) + \frac{i}{s} \left( \bfsfI - 3 \hat{\bf r}\otimes \hat{\bf r}\right) \right. \nonumber \\ 
 && -  \frac{1}{s^2} \left( \bfsfI - 3\hat{\bf r}\otimes \hat{\bf r}\right) \left.  \right],
\end{eqnarray}
again in terms of the dimensionless parameter $s = k_0 R$, analogous to Eq.~(\ref{eq:RWAScalarFieldKEquation2}) for scalar waves. The dyadic Green function~(\ref{eq:DyadicGreenFunctionFreeSpace}) has a near-field (NF) term that scales as $R^{-3}$, an intermediate-field (IF) term proportional to $R^{-2}$, and a far-field (FF) term $\propto R^{-1}$.
The second term in Eq.~(\ref{Eq:full_plus_error1}) for the RWA propagator is more interesting: it does not appear when using the full (i.e. non-RWA) light-matter interaction and this error term is again indeed purely real-valued (as in the scalar case), as expected from the discussion below Eq.~(\ref{eq:GisKminusK}).
We evaluate the derivatives also for this error term, which can be done using the identity for radial functions~\cite{Wubs:2015a}
\begin{equation}\label{eq:WubsDyadicDerivativeTrick}
    \nabla \otimes \nabla f(r) = \frac{f'(r)}{r}\left( \bfsfI - \hat{\bf r}\otimes \hat{\bf r}\right) + f''(r) \hat{\bf r}\otimes \hat{\bf r},
\end{equation}
and which for $f(R)=e^{-kR}/R$ leads  to
\begin{eqnarray}\label{eq:DyadicKGreenFunctionAndErrorRWA}
\mathbf{K}^{+}(R,k_0)/k_0 & = & \mathbf{G}_0(R,k_0)/k_0 - \frac{I_2(s)}{(2 \pi s)^2}\left(\bfsfI-\hat{\bf r}\otimes\hat{\bf r}\right)  \nonumber\\
-&& 
\frac{I_1(s)+I_{0}(s)}{(2 \pi s)^2}\left(\bfsfI-3\hat{\bf r}\otimes\hat{\bf r}\right).
\end{eqnarray}
Here, the family of integrals $I_n(s)$ is defined by
\begin{equation}
I_n(s)    \equiv \int_0^\infty \text{d}u \frac{u^n e^{-u}}{u^2 + s^2}.
\end{equation}
By considering how each term in Eq.~(\ref{eq:DyadicKGreenFunctionAndErrorRWA}) scales as a function of $R$, or equivalently of $s = k_0 R$, we can identify far-field (FF), intermediate-field (IF), and near-field (NF) terms also in the error term in Eq.~(\ref{eq:DyadicKGreenFunctionAndErrorRWA}).
We first consider the near-field limit ($s \to 0$), where we find that $I_0(s) \to \pi/(2 s)$ and $I_1(s) \to - \ln(s)$, and finally $I_2(s) \to 1$. 
Therefore, not only the free-space Green function but also the dominant RWA-error term scale as $R^{-3}$ in the near-field.
Their magnitudes differ appreciably, however: in the NF, the  real parts of the dyad in the two formalisms differ by a factor of two, since $\text{Re}\left[\mathbf{K}^{+}\right]/\text{Re}\left[\mathbf{G}_0\right] \to {1}/{2}$ as $R \to 0$. This limiting value of was also found in~\cite{FRIEDBERG:1973101}. This limiting behavior is in stark contrast to the scalar field case of Sec.~\ref{sec:TheScalarField}, where the error diverged faster than the scalar Green's function itself. 
The RWA dyad is smaller than the exact full dyad because the NF error term is smaller by a factor of two and has an opposite sign compared to the NF term of the free-space Green's function.

The $R \to 0$ limit of the near-field interaction is related to the single-atom self-interaction  $X$ of Eq.~(\ref{eq:RWAPoleShift}), where the two  positions in the propagator are both to be taken equal to the atomic position.   Formally the real part of this self-interaction  will diverge, as  for the scalar model in Sec.~\ref{sec:TheScalarField}.
This divergence can be taken care of by a renormalization~\cite{Bethe:1947a} or Green-function regularization~\cite{deVries1998a}. Agarwal~\cite{Agarwal:1971a,Agarwal:1973a} suggests that using the RWA in the study of the frequency shift of an atom corresponds to discarding the frequency shift of the lower level and including only the upper level shift, as also supported by Milonni and Knight~\cite{MilonniKnight:1974a}; Friedberg and Manassah argue that the RWA makes single-atom Lamb shifts a factor of 2 too small~\cite{FRIEDBERG:1973101}.
Here instead we will deal with this issue in the same pragmatic way as for the scalar case: we absorb into the bare transition frequency the free-space radiative shift via $\Omega + X\rightarrow \Omega^{\text{Obs}} + i\text{Im}\left[X\right]$, where $\Omega^{\text{Obs}}=\Omega+\text{Re}\left[X\right]$ is the observed transition frequency~\cite{Leonardi1986a,SokhoyanAtwater:2013}. Single-atom transition frequencies are then the same whether one makes the RWA or not. Henceforth all references to single-atom transition frequencies will be to the observable  transition frequencies, and so we will drop the superscript `Obs'. 

Going to the far field ($s \gg 1$) instead, to estimate the RWA error in the propagator, one can approximate the denominator $(u^2 + s^2)$ of $I_n(s)$ by $s^2$, which gives $I_n(s) \to n!/s^2$. The RWA error in the far field therefore falls off as $R^{-4}$, much faster than the term of the free-space Green function in Eq.~(\ref{eq:DyadicGreenFunctionFreeSpace}) proportional to $R^{-1}$ that dominates the far-field behavior. So no significant errors due to making the RWA will occur in the far field. 

We caution here that the results for the RWA error  were explicitly derived with the assumption that the coupling does not appreciably change the resonance frequency, i.e. by the pole approximation below Eq.~(\ref{eq:RWAPoleShift}). In particular one cannot take  the extreme near-field limit $k_0R\rightarrow0$ without violating the pole approximation at some point. Nevertheless the comparison of the near-field limits with and without the RWA is useful: a sneak preview of Fig.~\ref{fig:InteractionStrengthLimit2} shows that already for $k_0 R\approx1$, i.e. for distances for which the pole approximation is far from breaking down, the fraction of the real parts of the near-field interaction is already close to the discussed  $R \rightarrow 0$ limiting value of ${1}/{2}$. 
In Sec.~\ref{sec:2atomCollective_emission} we will 
give numerical examples of the error in the RWA electromagnetic propagator and its consequences.

\section{N-Atom Scattering}\label{Sec:Natom_scattering}
\noindent In the following, we will consider light emission by a multi-atom system using a scattering matrix formalism as in Ref.~\cite{Wubs2004a}, but unlike in that paper we will quantify the effect of making the RWA on collective emission.
We will adopt the similar shorthand notation of dropping function arguments and letting the subscript $m$ indicate that a function is to be evaluated at the coordinates of atom $m$. In the short-hand notation we also drop  the superscript $'+'$ for the moment.
In particular, we use the abbreviations  $\mathbf{F}_m=\mathbf{F}^{+}(\textbf{R}_m,\omega)$, $\mathbf{K}_m=\mathbf{K}^{+}({\bf r},\mathbf{R}_m,\omega)$, $\mathbf{K}_{mn}=\mathbf{K}^{+}(\mathbf{R}_m,\mathbf{R}_n,\omega)$ and we will introduce $\mathbf{F}^{(1)}=\mathbf{E}^{+(0)}(\mathbf{r},\omega) + \sum_{m}\mathbf{K}^{+}({\bf r},\mathbf{R}_m,\omega) \cdot\mathbf{S}^{+}(\mathbf{R}_m, \omega)$.
We can then iteratively write out Eq.~(\ref{eq:RWAFieldOperatorPlusMinusFreeSourceScattering}) and get 
\begin{eqnarray}
        \mathbf{F}-&&\mathbf{F}^{(1)} =  \sum_{n=1}^{N}\mathbf{K}_n \cdot \mathbf{V}_n \cdot \mathbf{F}_n^{(1)}\nonumber\\ + &&\sum_{m,n=1}^{N}\mathbf{K}_m \cdot \mathbf{V}_m \cdot \mathbf{K}_{mn} \cdot \mathbf{V}_n \cdot \mathbf{F}^{(1)}_n 
      \\  + &&\sum_{m,n,p=1}^{N}\mathbf{K}_m \cdot \mathbf{V}_m \cdot \mathbf{K}_{mp} \cdot \mathbf{V}_p \cdot \mathbf{K}_{pn} \cdot \mathbf{V}_n \cdot \mathbf{F}_n^{(1)} + ...\nonumber
\end{eqnarray}
At this point it pays off to introduce into this infinite series the single-atom scattering matrices $\mathbf{T}^{+}_n(\omega)$ (or $\mathbf{T}_n(\omega)$ in short-hand notation) that were defined in Eq.~(\ref{eq:SingleAtomScatteringMatrix}). 
These account for all sequences of repeated scattering any number of times off a single atom before propagation by the RWA propagator to another atom or to the observation point $\mathbf{r}$.
Doing so gives us
\begin{eqnarray}\label{eq:GeneralizedNatomLipmannSchwinger2}
\mathbf{F}-&&\mathbf{F}^{(1)}  = \sum_{n=1}^{N}\mathbf{K}_n \cdot \mathbf{T}_n \cdot \mathbf{F}_n^{(1)} \nonumber\\
+&& \sum_{m,n=1}^{N}\mathbf{K}_m \cdot \mathbf{T}_m \cdot \mathbf{K}'_{mn} \cdot \mathbf{T}_n \cdot \mathbf{F}^{(1)}_n  \\
        +&& \sum_{m,n,p=1}^{N}\mathbf{K}_m \cdot \mathbf{T}_m \cdot \mathbf{K}'_{mp} \cdot \mathbf{T}_p \cdot \mathbf{K}'_{pn} \cdot \mathbf{T}_n \cdot \mathbf{F}_n^{(1)} + ... ,\nonumber 
\end{eqnarray}
where $\mathbf{K}'_{mn}=(1-\delta_{m,n})\mathbf{K}_{mn}$ ensures propagation between different atoms. 
We can then write Eq.~(\ref{eq:GeneralizedNatomLipmannSchwinger2}) 
in a much more compact form as
\begin{eqnarray}\label{eq:GeneralizedNatomLipmannSchwinger3}
\mathbf{F}-\mathbf{F}^{(1)}  = \sum_{n,m=1}^N\textbf{K}_m\cdot \mathbf{T}^{(N)}_{mn} \cdot \mathbf{F}^{(1)}_n, 
\end{eqnarray}
where the generalized N-atom scattering matrix is given by
\begin{subequations}
\begin{equation} \label{eq:NAtomScatteringMatrix}
    \mathbf{T}^{(N)}_{mn}=\hat{\bm\mu}_m T_m M^{-1}_{mn} \hat{\bm\mu}_n,
\end{equation}
in terms of the $N \times N $ matrix
\begin{equation}\label{eq:NAtomScatteringMatrixMMatrix}
    M_{mn}=\delta_{mn} - (1-\delta_{mn})\hat{\bm\mu}_m \cdot \mathbf{K}_{mn} \cdot \hat{\bm\mu}_n T_n.
\end{equation}
\end{subequations}%
The $N$-atom scattering matrix in Eq.~(\ref{eq:NAtomScatteringMatrix}) takes into account any combination of scattering events among the $N$ atoms in the sample.
With these ingredients we can now write the full RWA field, dropping the short-hand notation,  as 
\begin{equation}
    \mathbf{F}^{+}(\mathbf{r},\omega)=\mathbf{E}^{+(0)}(\mathbf{r},\omega) + \mathbf{F}^{+}_{\rm source}(\mathbf{r},\omega) +\mathbf{F}^{+}_{\rm scat}(\mathbf{r},\omega),
\end{equation}
where 
\begin{equation}
    \mathbf{F}^{+}_{\text{scat}}(\mathbf{r},\omega)=\sum_{n,m=1}^N\textbf{K}^{+}(\mathbf{r}, \mathbf{R}_m, \omega)\cdot \mathbf{T}^{+(N)}_{mn}(\omega) \cdot \mathbf{E}^{+(0)}(\mathbf{R}_n, \omega)
\end{equation}
and 
\begin{equation}\label{eq:NAtomSourceTerm}
    \mathbf{F}^{+}_{\text{source}}(\mathbf{r},\omega)= \sum_{m=1}^N \mathbf{K}^{+(N)}(\mathbf{r}, \mathbf{R}_m, \omega)\cdot\mathbf{S}^{+}_m(\omega).
\end{equation}
Here we have introduced a generalized $N$-atom propagator
\begin{eqnarray} \label{eq:GeneralizedNAtomPropagator}
&& \mathbf{K}^{+(N)}(\mathbf{r}, \mathbf{r}', \omega) = \mathbf{K}^{+}(\mathbf{r}, \mathbf{r}', \omega)\\
&& + \sum_{m,n=1}^N \mathbf{K}^{+}(\mathbf{r}, \mathbf{R}_m, \omega) \cdot \mathbf{T}^{+(N)}_{mn}(\omega) \cdot \mathbf{K}^{+}(\mathbf{R}_n, \mathbf{r}', \omega).\nonumber
\end{eqnarray}
The interpretation is that while $\mathbf{K}^{+}(\mathbf{r}, \mathbf{r}', \omega)$ is the RWA propagator in the medium in the absence of the $N$ atoms, $ \mathbf{K}^{+(N)}(\mathbf{r}, \mathbf{r}', \omega)$ is the RWA propagator in the presence of these atoms. It is defined in terms of the $N$-atom T-matrix $\mathbf{T}^{+(N)}_{mn}(\omega)$, which  describes multiple light scattering up to infinite order among all atoms. 

\section{Two-atom Collective Emission in RWA}
\label{sec:2atomCollective_emission}
\subsection{The two-atom resonance condition} \label{sec:2atomResonanceCondition}
\noindent In order to quantify the effect of the RWA on superradiance, we first  consider the simplest case of two atoms, each with fixed dipole moments, separated along the z-axis by a distance $R$.
The two-atom version of Eq.~(\ref{eq:NAtomSourceTerm}) is given by
\begin{equation} \label{eq:2AtomSourceEquation}
    \mathbf{F}_{\text{source}}(\mathbf{r},\omega)=  \mathbf{K}^{(2)}(\mathbf{r}, \mathbf{R}_1, \omega)\cdot\mathbf{S}_1(\omega) + \mathbf{K}^{(2)}(\mathbf{r}, \mathbf{R}_2, \omega)\cdot\mathbf{S}_2(\omega).
\end{equation}
(Again we dropped  RWA superscripts $'+'$ for readability.) We will focus on the resonance condition for just one of the two terms, and to continue we need to calculate the generalized two-atom propagator.
Considering now the two-atom $\mathbf{M}$-matrix of Eq.~(\ref{eq:NAtomScatteringMatrixMMatrix}), we see that it has the form
\begin{eqnarray}\label{eq:2AtomMMatrixEquation}
\mathbf{M}(\omega)=
       \begin{pmatrix}
    1 & -J_{12} T_2/\beta \\
    -J_{21}T_1/\beta  & 1
    \end{pmatrix},
\end{eqnarray}
where we have introduced a short-hand notation $\beta \equiv \mu_m^2 \omega^2/(\hbar \epsilon_0 c^2)$, and we defined the interatomic interaction  
\begin{equation}\label{Eq:interatomic_interaction}
    J_{12}=\beta \hat{\bm{\mu}}_1 \cdot \bm{K}_{12}\cdot \hat{\bm{\mu}}_2.
\end{equation}
From Eq.~(\ref{eq:DyadicKGreenFunctionAndErrorRWA}) it is clear that making the RWA changes this two-atom interaction. 
Inverting the matrix in Eq.~(\ref{eq:2AtomMMatrixEquation}) and inserting this into Eq.~(\ref{eq:NAtomScatteringMatrix}) yields
\begin{equation}\label{eq:2AtomScatteringMatrix}
\textbf{T}^{(2)}=\frac{1/\beta}{ 1- T_1 J_{12}^2T_2/\beta^2}
    \begin{pmatrix}
    \hat{\bm{\mu}}_1 \hat{\bm{\mu}}_1 \beta T_1 & \hat{\bm{\mu}}_1 \hat{\bm{\mu}}_2 T_1 J_{12} T_2 \\
    \hat{\bm{\mu}}_2 \hat{\bm{\mu}}_1 T_2 J_{12} T_1 & \hat{\bm{\mu}}_2 \hat{\bm{\mu}}_2 \beta T_2
    \end{pmatrix}.
\end{equation}
We will now adopt the short-hand notation of $\mathbf{K}(\mathbf{r}1)\equiv\mathbf{K}(\mathbf{r},\mathbf{R}_1)$ and $\mathbf{K}(11)\equiv\mathbf{K}(\mathbf{R}_1,\mathbf{R}_1)$ and so on, to shorten the notational complexity.
We can write out the generalized propagator for the first atom via Eqs.~(\ref{eq:GeneralizedNAtomPropagator}) and (\ref{eq:2AtomScatteringMatrix}) as
\begin{eqnarray}
    \mathbf{K}^{(2)}(\mathbf{r}1) 
      = &&\mathbf{K}(\mathbf{r}1)\cdot \left[ \bfsfI + \mathbf{T}_{11} \cdot \mathbf{K}(11) +  \mathbf{T}_{12} \cdot \mathbf{K}(21)\right] \nonumber\\
      + &&\mathbf{K}(\mathbf{r}2)\cdot \left[ \mathbf{T}_{21}  \cdot \mathbf{K}(11) + \mathbf{T}_{22} \cdot \mathbf{K}(21)\right].
\end{eqnarray}
From here, it is straightforward to write out the source term resulting from the first atom in Eq.~(\ref{eq:2AtomSourceEquation}),
\begin{eqnarray} \label{eq:2AtomResonanceEquation1}
        \mathbf{K}^{(2)}(\mathbf{r}1) \cdot &&\mathbf{S}_1 = \left(\frac{1 + T_1 X_1/\beta}{1- T_1 J_{12}^2T_2/\beta^2}\right)\nonumber \\
        \times&&\left[\mathbf{K}(\mathbf{r}1)\cdot\hat{\bm{\mu}}_1 +\mathbf{K}(\mathbf{r}2)\cdot\hat{\bm{\mu}}_2 T_2 J_{12}/\beta  \right]S_1.
\end{eqnarray}
Inserting the definitions from Eqs.~(\ref{eq:RWADyadAndVectorDefinitions}) we arrive at 
\begin{widetext}
\begin{equation}\label{eq:2AtomResonanceEquation2}
    \mathbf{K}^{(2)}(\mathbf{r}1) \cdot \mathbf{S}_1 = \frac{(\omega- \Omega_1)S_1\left[ \mathbf{K}(\mathbf{r}, \mathbf{R}_1) \cdot \hat{\bm{\mu}}_1 (\omega - \Omega_2 - X_2) +   \mathbf{K}(\mathbf{r}, \mathbf{R}_2) \cdot \hat{\bm{\mu}}_2 J_{12}\right]}{(\omega - \Omega_1 -X_1)(\omega - \Omega_2 -X_2) - J_{12}^2}.
\end{equation}
\end{widetext}
The  source term is resonant for the zeroes of the denominator of Eq.~(\ref{eq:2AtomResonanceEquation2}).
We will once again make use of the pole approximation and assume that $X_n$ and $J_{12}$ do not change appreciably between being evaluated at the bare or the dressed transition frequencies. 
It then easily follows that two resonances exist, at 
\begin{eqnarray} \label{eq:2AtomSuperradiantResonanceFrequencies}
    \omega^{(\pm)}=&&\frac{\Omega_1  + \Omega_2}{2} + \frac{X_1 + X_2}{2} \\
    \pm &&\sqrt{\left(\frac{\Omega_1 + X_1 -\Omega_2 - X_2}{2}\right)^2 +J_{12}^2}\nonumber.
\end{eqnarray}
This expression is symbolically identical to the one derived with the full Hamiltonian in Ref.~\cite{Wubs2004a}. Numerically there will however in general be differences when making the RWA.

In the remainder of this section,  we will study the effect of making the RWA on the interatomic interaction~(\ref{Eq:interatomic_interaction}) in Sec.~\ref{sec:relativeError}, on the collective decay rates for two identical atoms in Sec.~\ref{Sec:2identical}, and on the collective decay rates for two detuned atoms in Sec.~\ref{Sec:two_detuned}.

\subsection{Effect of the RWA on interatomic interactions}\label{sec:relativeError}
\begin{figure*}[t!] 
    \centering
    \begin{subfigure}[b]{0.5\textwidth}
        \centering   \includegraphics[height=2in]{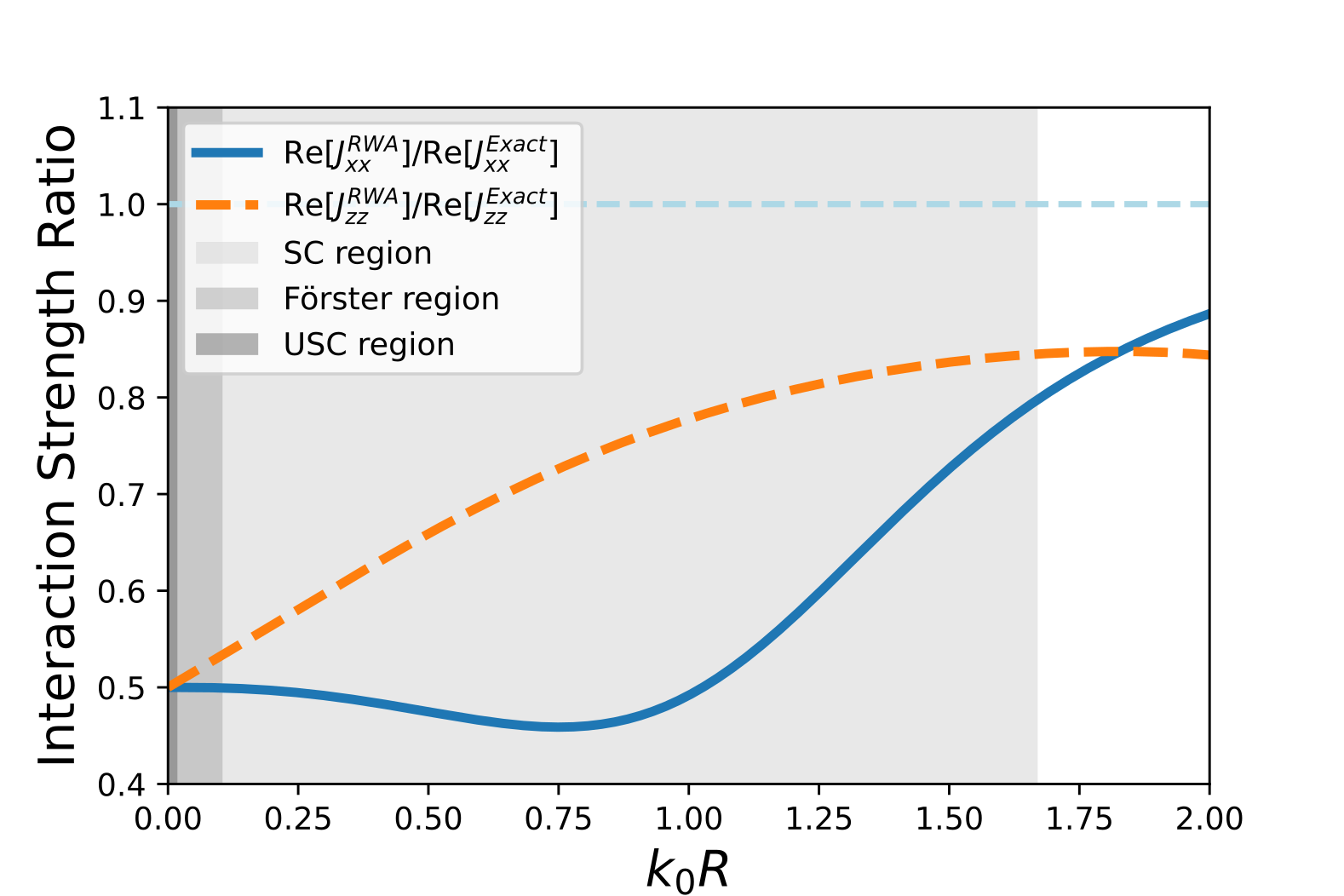}
        \caption{}
        \label{fig:InteractionStrengthLimit2}
    \end{subfigure}%
    \begin{subfigure}[b]{0.5\textwidth}
        \centering
        \includegraphics[height=2in]{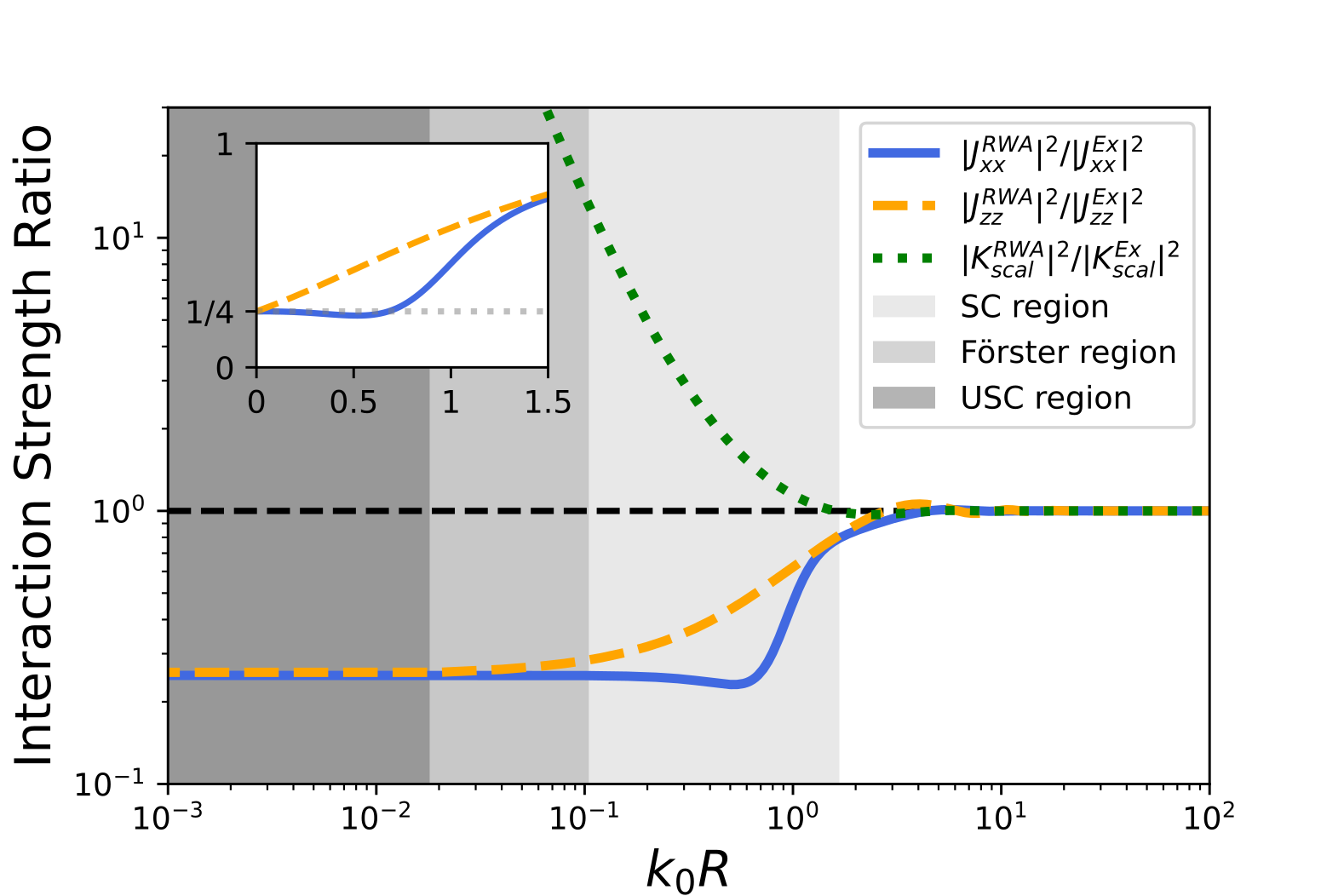}
        \caption{}
        \label{fig:InteractionStrengthLimit2LogLog}
    \end{subfigure}
    \caption{(a) Ratio of the real parts of the interatomic interaction Eq.~(\ref{Eq:interatomic_interaction}) with and without the RWA, for the x-x- and z-z-configurations. The grey regions are based on the z-z-configuration.
    (b) Ratio of the magnitudes squared of the interaction term.
    The grey shades, from light to dark, represent the regions of strong coupling, Förster transfer and ultra-strong coupling. The inset shows the same ratio on linear axes. The limiting ratio for $R \to 0$ is $\left(1/2\right)^2$.
    For the x-x configuration the ratio even assumes values smaller than 0.25 for finite distances around $k_0 R \simeq 0.8$.}
    \label{fig:InteractionStrengthLimit2Scalar}
\end{figure*}

\noindent We now turn our attention to  the interatomic interaction~(\ref{Eq:interatomic_interaction})  when utilizing the RWA. Our main question is whether making the RWA has negligible influence on this interaction or not. To be concrete, we will consider two  atomic  constellations, both with parallel dipoles: in the $x-x$ constellation, the atomic dipole moments are aligned in parallel and are orthogonal to the axis of separation ($z$-axis), while in the $z-z$ constellation, the atomic dipole moments and the separation axis all are parallel. It is known and we can see from Eq.~(\ref{eq:DyadicGreenFunctionFreeSpace}) that these different dipole constellations interact very differently, since $\hat{\bm \mu}\cdot\hat {\bf r}$ equals unity for the one and zero for the other case. In particular, the far-field term cancels out completely for the $z-z$ constellation. 
In Fig.~\ref{fig:InteractionStrengthLimit2Scalar} 
we visualize the effect of the RWA on the interaction as a function of the interatomic separation, on a linear distance scale in panel~(a) and on a logarithmic scale in the complementary panel~(b), for the $x-x$ and the $z-z$ configurations, and in (b) also for the scalar model. 

As the distance is decreased in Fig.~\ref{fig:InteractionStrengthLimit2Scalar}, we go through various regimes of typical interatomic separations that we define in detail below: starting with the largest distances, we have the   weak-coupling regime, then the strong-coupling region, followed by what we call the F\"orster regime, and finally the regime of ultrastrong coupling. We will not discuss the physics in all of these regimes, 
but the shaded regions are merely intended to give an overview of the effects of the RWA on interatomic interactions on typical length scales. The interatomic interaction~(\ref{Eq:interatomic_interaction}) is complex-valued. In panel~\ref{fig:InteractionStrengthLimit2Scalar}(a) we depict the ratio of the real parts of the RWA and full interatomic interactions, while in panel~\ref{fig:InteractionStrengthLimit2Scalar}(b) we plot the fraction of their absolute values squared $|J_{12}^{\rm RWA}/J_{12}|^2$. 


The {\em weak-coupling  region}  is unshaded in Fig.~\ref{fig:InteractionStrengthLimit2Scalar}. It is the regime where the magnitude of the interaction strength is smaller than the single-atom decay rate.  It can be seen especially in panel~(b) that the RWA and full interatomic interactions tend to the same values in this weak-coupling region, becoming almost equal in the far field ($k_0 R \gg 1$). The far field is part of the weak-coupling region, and the RWA is a good approximation in the far field, as expected. 
However, differences between the real parts of the RWA and full interatomic interactions can get as large as twenty percent even in this weak-coupling region, as panel~\ref{fig:InteractionStrengthLimit2Scalar}(a) illustrates. 


Next, the {\em strong-coupling  region} of Fig.~\ref{fig:InteractionStrengthLimit2Scalar} is lightly shaded in both panels. We use the common definition of the onset of strong coupling to be when the magnitude of the interaction strength is equal to the single-atom decay rate, i.e. when $\left|\text{Re}\left[J_{zz}\right]\right|\geq-\text{Im}\left[X\right]$ in our notation. The lightly shaded region corresponds in particular to  two dipoles in the z-z configuration,  which are strongly coupled for $k_0R\leq1.67$ or $R \le 0.27 \lambda$. 
Likewise for the x-x-configuration, the strong-coupling criterion holds for $k_0R\leq1.10$ or $R \le 0.18 \lambda$. 
This strong-coupling criterion invites an analogy with cavity quantum electrodynamics (CQED), where the strong-coupling regime is defined by a similar condition~\cite{NovotnyHecht:2000a,Black:2005a}. 
In CQED the cavity loss rate competes with the light-matter interaction, and strong coupling is defined as when the chance of a cavity mode being reabsorbed exceeds the chance that it will be lost out of the cavity.
In our analogy, strong coupling is achieved when the chance of the atomic excitation 'jumping' to the other atom exceeds the chance that the excitation will be lost by single-atom spontaneous decay. 
It is clear from Fig.~\ref{fig:InteractionStrengthLimit2Scalar} that the RWA and full interatomic interactions start to deviate considerably in the strong-coupling regime. The scalar model shows the largest deviations, with the scalar RWA interaction becoming more than a factor of ten too large when decreasing distances  within this regime. But also the dyadic interatomic interactions are affected, with the fraction of RWA and full interaction strengths tending towards the limiting value of $0.5$ already within this strong-coupling regime. So the RWA is not a good approximation in this strong-coupling regime, and one expects to see effects of this below when calculating collective emission rates in the RWA. Let us also mention that atoms in a gas at room temperature and at standard pressure have neighbors at an average distance of around 5~nm, with which they interact strongly (but briefly due to their thermal motion).


As the third interaction regime we identify the {\em F{\"o}rster regime}, denoted by medium grey shading in Fig.~\ref{fig:InteractionStrengthLimit2Scalar}.  F{\"o}rster resonance energy transfer (FRET)~\cite{Foerster1948a,Andrew:2000a,Wubs_2016} in a homogeneous medium occurs at the characteristic rate proportional to $R^{-6}$, with a typical separation $R$  between a donor and an acceptor of 3 to 10~nm, for experiments with dyes or quantum dots~\cite{Andrew:2000a,Blum:2012a,Xia:2019a}.
For a separation of 10 nm~\cite{Blum:2012a} and an emission wavelength of 600~nm, we find that this regime starts at $k_0 R=0.10$,  and ends at $k_0 R=0.03$ for $R = 3$~nm. Within the entire F{\"o}rster regime, the RWA and full interatomic interaction strengths can be seen to differ approximately by the limiting factor of  2, i.e the strength ratio is 0.5. The RWA should thus not be used to calculate F{\"o}rster energy transfer rates, otherwise one would find rates that correctly  scale as $R^{-6}$, yet are a factor of four too small.


Finally there is the {\em ultrastrong-coupling regime}, denoted by the darkest grey shading in Fig.~\ref{fig:InteractionStrengthLimit2Scalar}. The onset of this regime is defined by ultra-strong coupling (USC) criterion, being the point at which the interaction strength exceeds 10\% of the transition frequency~\cite{ciuti:2005a,Fuchs:2009a,FornDiaz:2010,Solano2010a}, or
$\text{Re}\left[J_{ii}\right]\geq 0.1 \Omega$ in our notation. For a quantum dot of dipole moment $\mu=9.7 \cdot 10^{-29}\text{C}\cdot\text{m}$ \cite{Stobbe20074387029} and transition frequency $\Omega=2\pi\cdot 500 $ $\text{THz}$ (corresponding to $\lambda=600$ nm),  we find that this threshold is reached for $J_{zz}$ when $k_0R=0.018$ or $R \le 0.003 \lambda$. 
It is within this regime that the pole approximation used in Secs.~\ref{sec:TheAtomicSourceTerm} and \ref{sec:2atomResonanceCondition} breaks down, and actually at these 2~nm or smaller distances the point-dipole approximation of a quantum dot  is also far from being valid.

Ref.~\cite{ciuti:2005a} states that 
the relative importance of the antiresonant
terms in the light-matter coupling is negligible if no
strong driving field is present, but important 
in the presence
of a strong driving field, giving rise to the Bloch-Siegert shift~\cite{BlochSiegert:1940a,Allen:1975}. 
By contrast, our results presented in Fig.~(\ref{fig:InteractionStrengthLimit2Scalar}) show that significant deviations due to making the RWA can be present even under weak driving, if only one has {\em two} emitters close enough to be in the strong-coupling regime.

\subsection{Collective emission by two identical atoms}\label{Sec:2identical}

\noindent Now that we have discussed the effect of the RWA on interatomic interactions, we investigate how the RWA affects collective emission rates. We start with the simplest case: for two identical atoms $(\Omega_1 = \Omega_2 = \Omega)$ experiencing the same effects of the medium $(X_1 = X_2 = X)$,  Eq.~(\ref{eq:2AtomSuperradiantResonanceFrequencies}) reduces to
\begin{equation}\label{eq:IdenticalAtomSuperradiantMode}
 \omega^{(\pm)}=\Omega + X \pm J_{12}.   
\end{equation}
for the two-atom resonances, with collective decay rates  
\begin{equation}\label{eq:TwoAtomIdenticalColectiveDecayRate}
    \gamma^\pm=-\text{Im}\left[X \pm J_{12}\right]=\gamma \mp i\,\text{Im}\left[J_{12}\right],
\end{equation}
where $\gamma=-\text{Im}\left[X(\Omega)\right]$ is again the single-atom decay rate.
If the atoms have parallel dipole orientations, then in the limit of $k_0 R\rightarrow0$ the interaction term approaches the self-interaction, i.e. $J_{12}\rightarrow X$. This result is valid whether one makes the RWA or not. 
The collective decay rates of the two very closely spaced emitters are then zero and twice the single-atom decay rate, corresponding to sub- and superradiance, respectively. 

So what are the effects of the RWA on collective light emission? Recall that the error term in $\mathbf{K}_{\rm RWA}$ in Eq.~(\ref{eq:DyadicKGreenFunctionAndErrorRWA}) is purely real-valued, and so is the RWA error in the interatomic interaction $J_{12}$ of Eq.~(\ref{Eq:interatomic_interaction}). We therefore find as an important result that {\em the super- and subradiant decay rates of two identical atoms in a nondispersive non-absorbing medium are unaffected by making the RWA, whatever their interatomic distance}. This result is similar to the single-atom spontaneus-emission rate, which in Sec.~\ref{sec:TheAtomicSourceTerm} was also shown to be unaffected by making the RWA.  

By contrast, making the RWA does have an effect on the two-atom sub- and superradiant resonance frequencies: the superradiant resonance frequency equals $Re[\omega^{(+)}]= \Omega + Re(J_{12})$,  and the subradiant frequency $Re[\omega^{(-)}]= \Omega - Re(J_{12})$. These frequencies and their splitting of $2 |Re(J_{12})|$ are affected by the RWA, because of the effect of the RWA on $Re(J_{12})$ as we discussed in Sec.~\ref{sec:relativeError}.  

\begin{figure*}[t!] 
    \centering
    \begin{subfigure}[b]{0.5\textwidth}
         \centering   \includegraphics[height=2in]{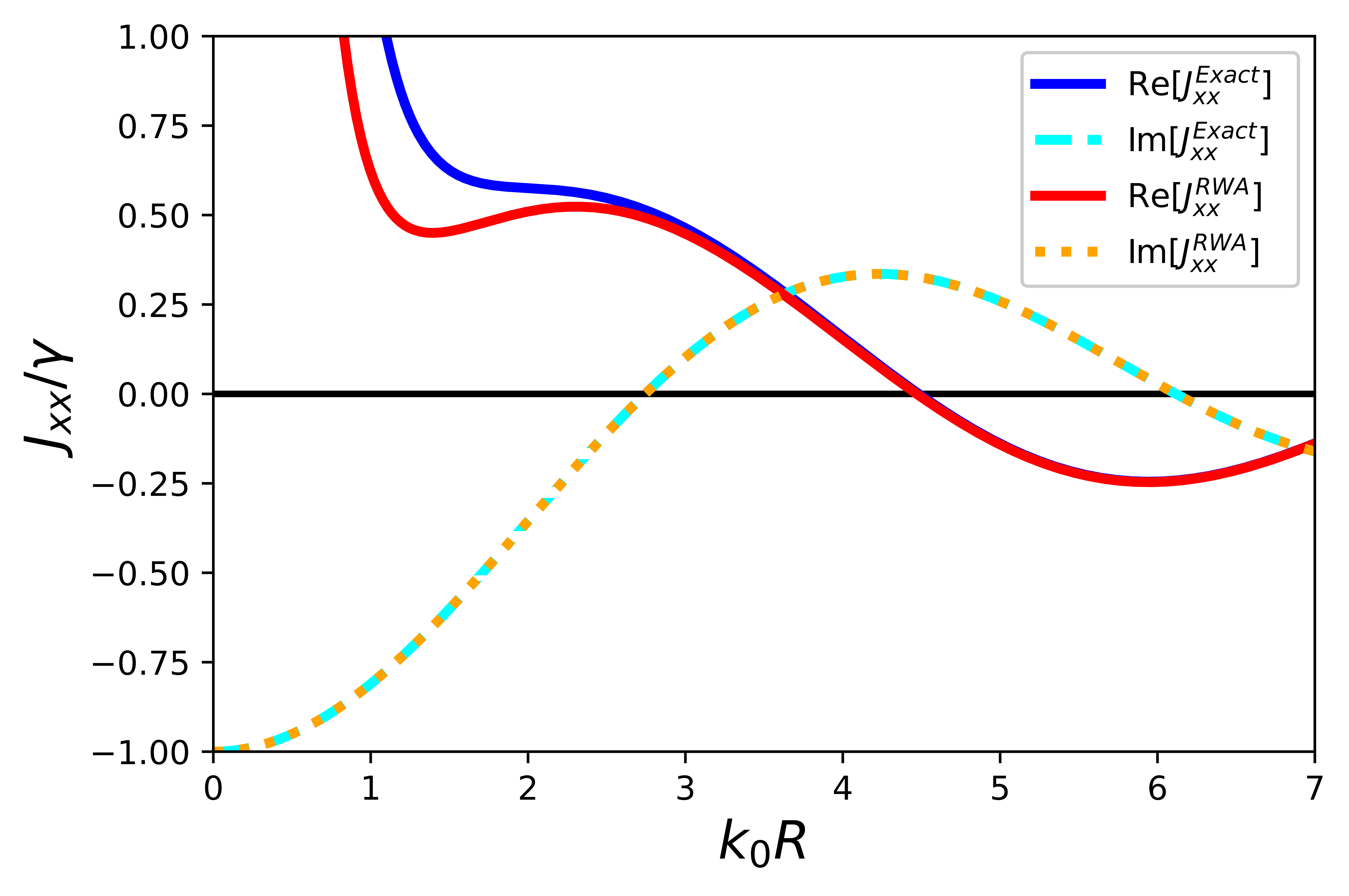}
        \caption{}
        \label{fig:JxxReImRWAExact}
    \end{subfigure}%
    \begin{subfigure}[b]{0.5\textwidth}
        \centering
        \includegraphics[height=2in]{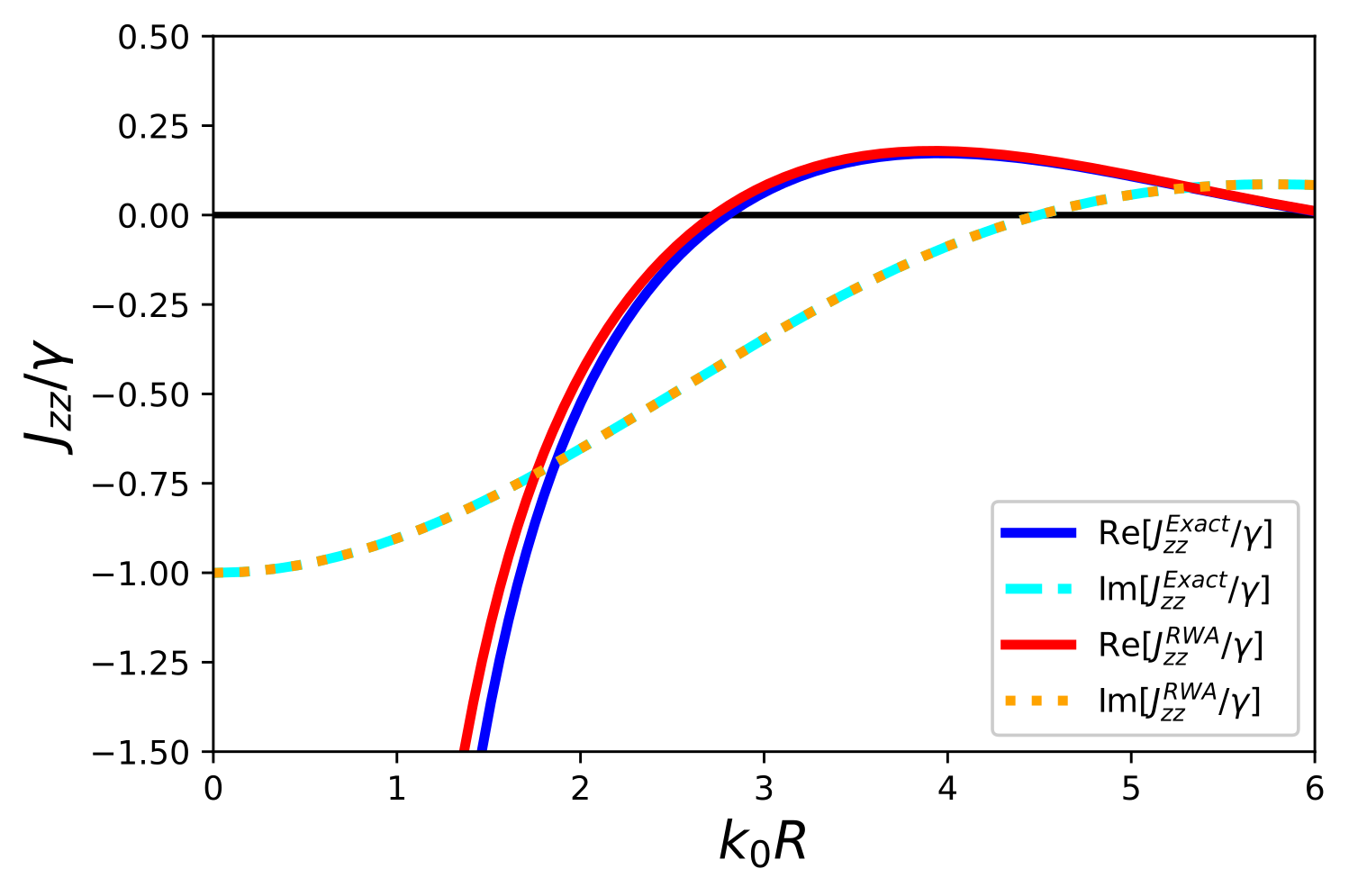}
        \caption{}
        \label{fig:JzzReImRWAExact}
    \end{subfigure}
    \caption{Real and imaginary parts of the interatomic interaction Eq.~(\ref{Eq:interatomic_interaction}), scaled by the single-atom spontaneous-decay rate $\gamma$, for the two constellations described in the main text. Panel~(a): $J_{xx}$, and (b): $J_{zz}$. The imaginary parts of the RWA and exact results are identical, as the RWA error of the interatomic interaction is purely real-valued.}
    \label{fig:JxxJzzReImRWAExact}
\end{figure*}

In Fig.~\ref{fig:JxxJzzReImRWAExact} 
we show the real and imaginary parts of the interatomic interaction of Eq.~(\ref{Eq:interatomic_interaction})  in the z-z and x-x constellations. In panel~(a) we see that the same $Im(J_{12})$ is found for the $x$--$x$ configuration, whether one makes the RWA or not.   The same holds true for the $z$--$z$ configuration in panel~(b). This illustrates that for two identical atoms, the two predicted collective decay rates indeed do not depend on making the RWA or not. By contrast,  $Re(J_{12})$ and hence the level splitting $Re[\omega^{(+)}-\omega^{(-)}]$ in both panels does depend on the RWA, with differences increasing with decreasing interatomic distances. 
The RWA error in the predicted level splitting becomes noticeable for  $k_0 R \le 2$.       

One can define a Rayleigh criterion for visibility of collective emission in the spectrum, equivalent to the Rayleigh criterion for the double-peaked spectrum characteristic of Rabi oscillations~\cite{Thompson:1992a}. From Eq.~(\ref{eq:IdenticalAtomSuperradiantMode}) it follows that the super- and subradiant emission peaks are separated by $\text{Re}\left[\omega^{(+)}-\omega^{(-)}\right]=2\text{Re}\left[J_{12}\right]$. The widths of the spectral peaks, given by twice the decay rate, are different for the super- and subradiant modes.
If we use the sum of half the widths of each peak to compare with the peak separation, we get $-\text{Im}\left[\omega^{(+)}+\omega^{(-)}\right]=2\gamma$. The Rayleigh criterion therefore becomes $\text{Re}\left[J_{zz}\right]\geq\gamma$, identical to the   strong-coupling criterion that we used in  Sec.~\ref{sec:relativeError}. The left-hand side of this inequality is affected by making the RWA, while the right-hand side is not. Likewise one can define the  cooperativity, $C=(\gamma_c - \gamma)/\gamma$, where $\gamma_c=-\text{Im}\left[\omega\right]$ is the cooperative decay rate of the system, quantifies the deviation from the single-atom decay rate~\cite{Leymann:2015a}. The cooperativity is not affected by making the RWA, not for two identical emitters at least.

The take-home message of this subsection seems to be that predicted collective decay rates are unaffected by making the RWA, while the collective resonance energies are affected.  
While true for two identical atoms, this message cannot be generalized, however. First, it does not imply that collective decay rates of two non-identical atoms can be computed accurately within the RWA (see Sec.~\ref{Sec:two_detuned} below). And secondly, unlike for two atoms, the collective decay rates of three or more identical atoms will in general be affected by making the RWA, as we will see in  Sec.~\ref{Sec:three_RWA} below.

\subsection{Collective emission by two detuned atoms }\label{Sec:two_detuned}

\noindent If the two atoms differ, for example through a detuning that is small compared to the transition frequencies, then their collective emission is slightly more complex than described in the previous section~\ref{Sec:2identical}, and the effect of the RWA is qualitatively different, as we shall see here.
In many practical applications the emitters (here called atoms) 
in fact are artificial atoms such as quantum dots or fluorescent colour centres.
Such solid-state quantum emitters are likely to exhibit inhomogeneous broadening, in other words their optical transition frequencies will differ due to variations in fabrication conditions like size and strain, as well as changes due to the local photonic environment.
It is thus relevant to study the effect of the RWA  on such slightly detuned two-atom systems, and again we will focus on their collective emission rates. For simplicity we do not take non-radiative decay rates into account, which is an idealization for most solid-state emitters. Superradiance in inhomogeneous ensembles, including detuning, has recently been studied in Refs.~\cite{Damanet:2016a,andrejicpalffy2021a}. 
Here we focus on the effect of the RWA for two slightly detuned atoms in free space. Given two single-atom transition frequencies $\Omega_1$ and $\Omega_2$, we introduce their detuning $\delta=(\Omega_1 - \Omega_2)/2$ and central frequency $\Omega=(\Omega_1 + \Omega_2)/2$.
For simplicity we will consider emitters with identical dipole moment magnitudes and only a tiny detuning compared to the observable transition frequencies. 
For free space the interaction term $J_{12}$ can then be evaluated at the central frequency and also the two single-atom spontaneous-emission rates are well approximated by the single-atom free-space decay rate
$-\text{Im}\left[X(\Omega)\right]\equiv  \gamma$.
The pole approximation, together with this assumption, hold for detunings up to multiple orders of magnitude of $\gamma$. For example, for $\lambda=600$ nm corresponding to $\Omega=2\pi \cdot500$ THz$\approx 3\cdot 10^{15} $ $\text{s}^{-1}$, we have $\gamma\approx 6.1\cdot 10^{8}$ $\text{s}^{-1}$. Therefore, a detuning even up to $\delta/\gamma\approx 10^4$ does not appreciably change the single-atom emission properties described by $X_n(\Omega_n)$.

Having thus argued that the pole approximation is still valid for the detuned two-atom system, we find from Eq.~(\ref{eq:2AtomSuperradiantResonanceFrequencies}) that the complex frequencies for collective emission  then become
\begin{eqnarray} \label{eq:2AtomDetunedSuperradiantResonanceFrequencies}
    \omega^{(\pm)}=\Omega - i\gamma
    \pm \sqrt{\delta^2 +J_{12}^2},
\end{eqnarray}
with detuning-dependent decay rates
\begin{eqnarray}\label{eq:2AtomDetudenSuperradianceDecayRate}
\gamma_\delta^\pm= \gamma \mp \text{Im}\sqrt{\delta^2+J_{12}^2}.
\end{eqnarray}
From this it is clear that some of the purely real-valued RWA error of $J_{12}$ will `spill over' into the imaginary part of the resonances, $\text{Im}\left[ \omega^{(\pm)}\right]$, and thereby affect the collective decay rates $\gamma_\delta^\pm$, in contrast to the case for identical emitters considered previously. 
The errors in the collective decay rates are purely due to the error incurred in the propagator when applying the Rotating-Wave Approximation.

\begin{figure*}[t!] 
    \centering
    \begin{subfigure}[b]{0.5\textwidth}
        \centering
          \includegraphics[height=2in]{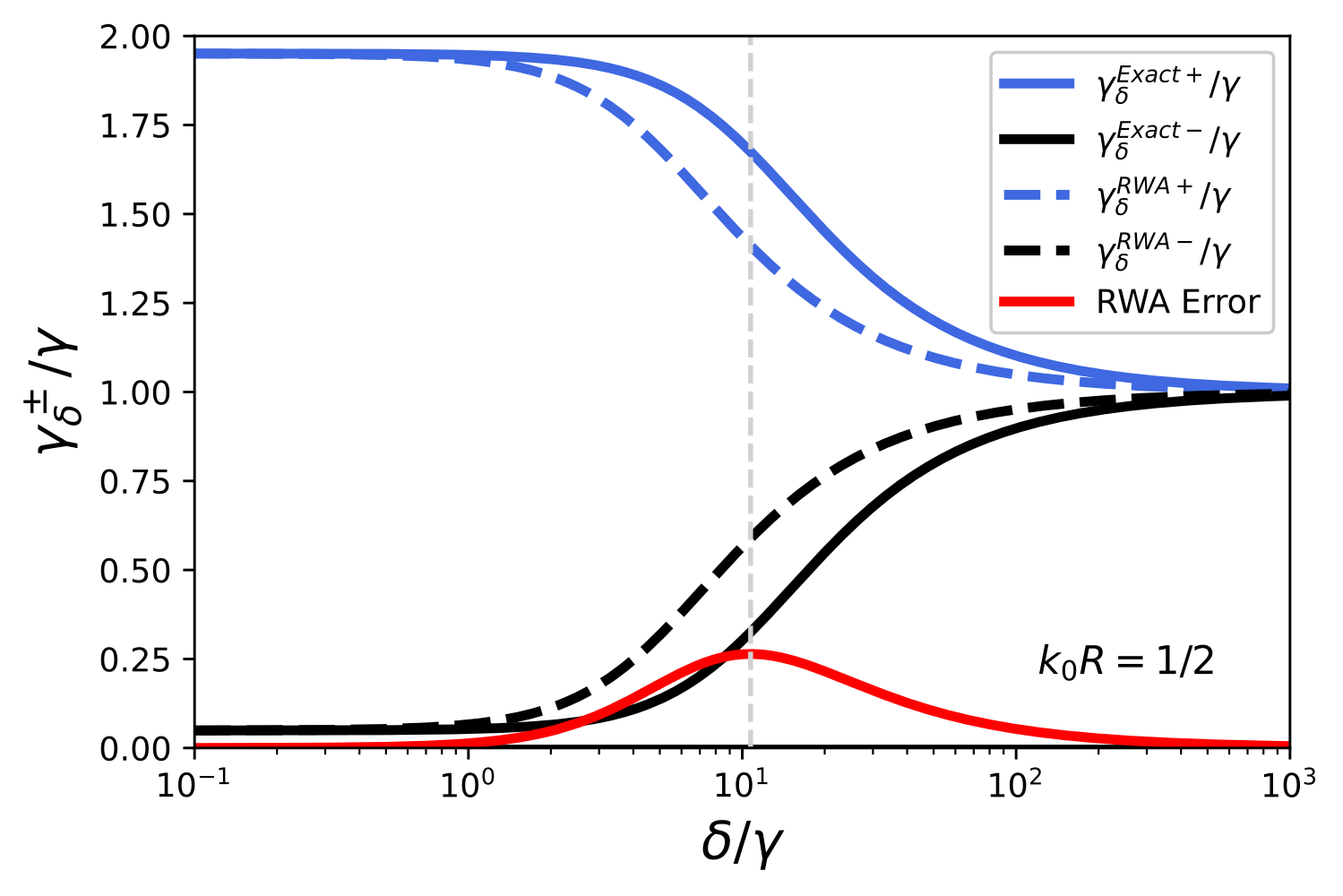}
        \caption{}
        \label{fig:2AtomDetunedResonanceDetuningDependence}
    \end{subfigure}%
    \begin{subfigure}[b]{0.5\textwidth}
        \centering
         \includegraphics[height=2in]{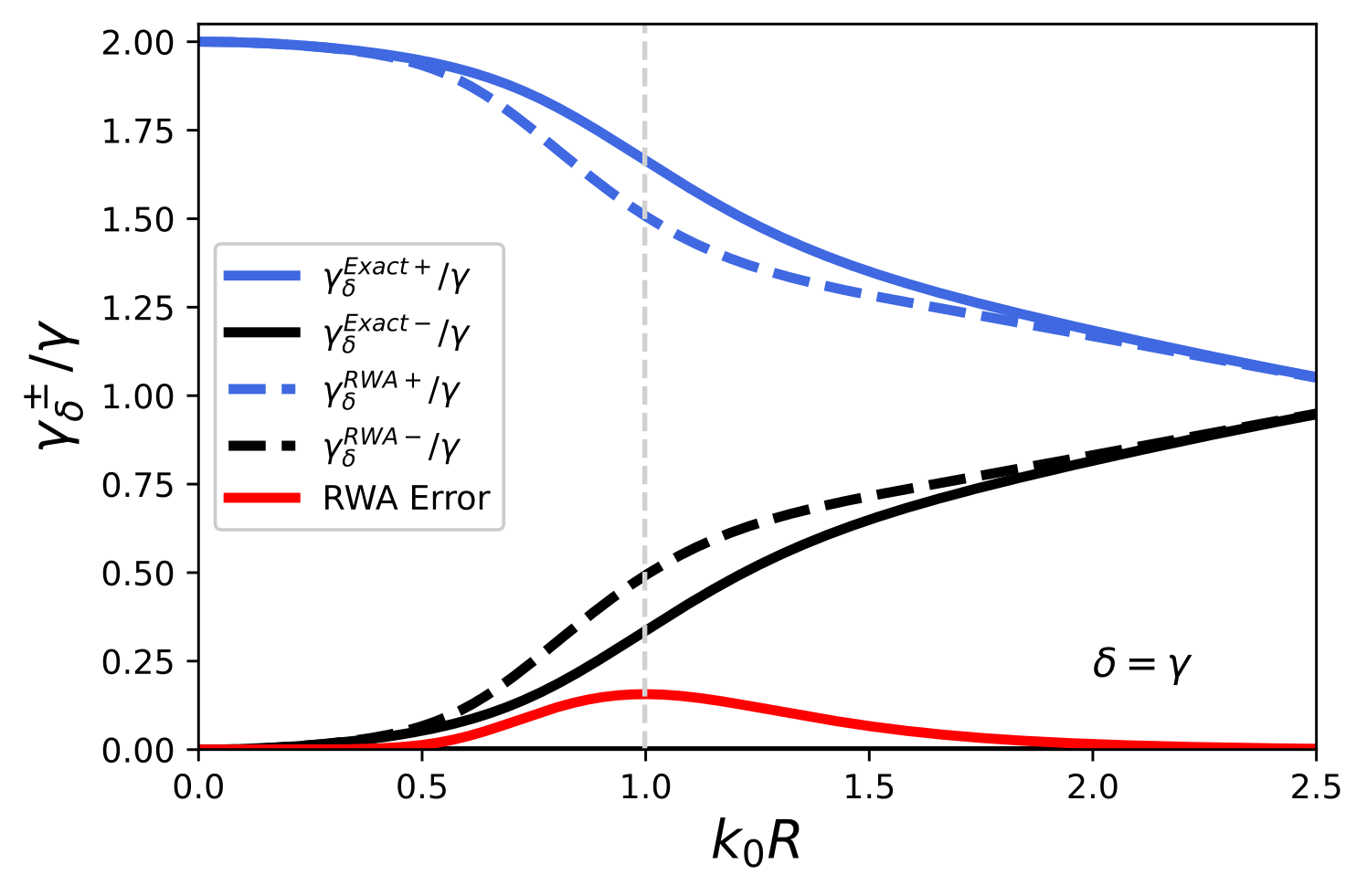}
        \caption{}
        \label{fig:2AtomDetunedResonanceSeparationDependence}
    \end{subfigure}
    \caption{The collective decay rates of two atoms in the RWA (dashed lines) as compared to full-interaction (solid lines) formalisms. (a) Detuning dependence of two detuned atoms at a distance  given by $k_0R =1$. 
    (b) Super- and subradiant decay rates at a finite detuning $\delta=2\gamma$, as a function of interatomic separation. In both panels, the maximal error due to making the RWA is indicated by a vertical dashed line. 
    }
    \label{fig:2AtomDetunedResonance}
\end{figure*}

Fig.~\ref{fig:2AtomDetunedResonanceDetuningDependence}
shows the decay rates of the collective resonance of an x-x configuration in the exact and the RWA formalisms for a fixed separation as a function of the detuning. Vice versa, Fig.~\ref{fig:2AtomDetunedResonanceSeparationDependence} shows these collective resonances as a function of separation for a fixed detuning.
The panels  show how detuning competes with the interaction strength, the one quenching and the other enhancing  collective emission. 
The collective contribution to the decay rate is diminished with increasing $\delta$, or by decreasing the interaction strength by increasing the inter-atomic distance. 
In the limits of high detuning or large separation, the two collective decay rates  become degenerate again and  equal to the single-emitter decay rate $\gamma$, as expected for two well-isolated atoms.

Interestingly, in both panels of Fig.~\ref{fig:2AtomDetunedResonance} we see that the RWA and exact decay rates for two detuned atoms only agree in the extreme cases: in panel~(a) both for vanishing detuning (ideal for collective emission) and for large detuning (emission no longer collective); analogously in panel~(b) both for vanishing distance and for large distances.  In between these extreme cases, there is a finite error in the collective decay rates due to making the RWA. 

In fact, for a fixed separation $R$, there must be some finite detuning $\delta$ (and vice versa) for which the error in the predicted decay rates is largest. The maximum discrepancy is found by comparison of Eq.~(\ref{eq:2AtomDetudenSuperradianceDecayRate}) in the RWA and exact formalisms.
The resulting  relative RWA error in the collective decay rates is shown in Fig.~\ref{fig:2AtomDetunedResonance}.  
Interestingly, the maximal relative error of even around $25 \%$ in panel~(a) occurs for  $\delta/\gamma$ of order unity, and in panel~(b) for a value of $k_0 R$ close to unity. So here we find considerable effects of making the RWA on predicted collective decay rates of slightly detuned emitters, occurring neither in the near- nor in the far-field, but rather in the  intermediate regime of distances of the order of the optical wavelength.

\section{Three-atom emission in RWA}\label{Sec:three_RWA}

\begin{figure*}[t!] 
    \centering
    \begin{subfigure}[b]{0.25\textwidth}
        \centering          \includegraphics[height=1.6in]{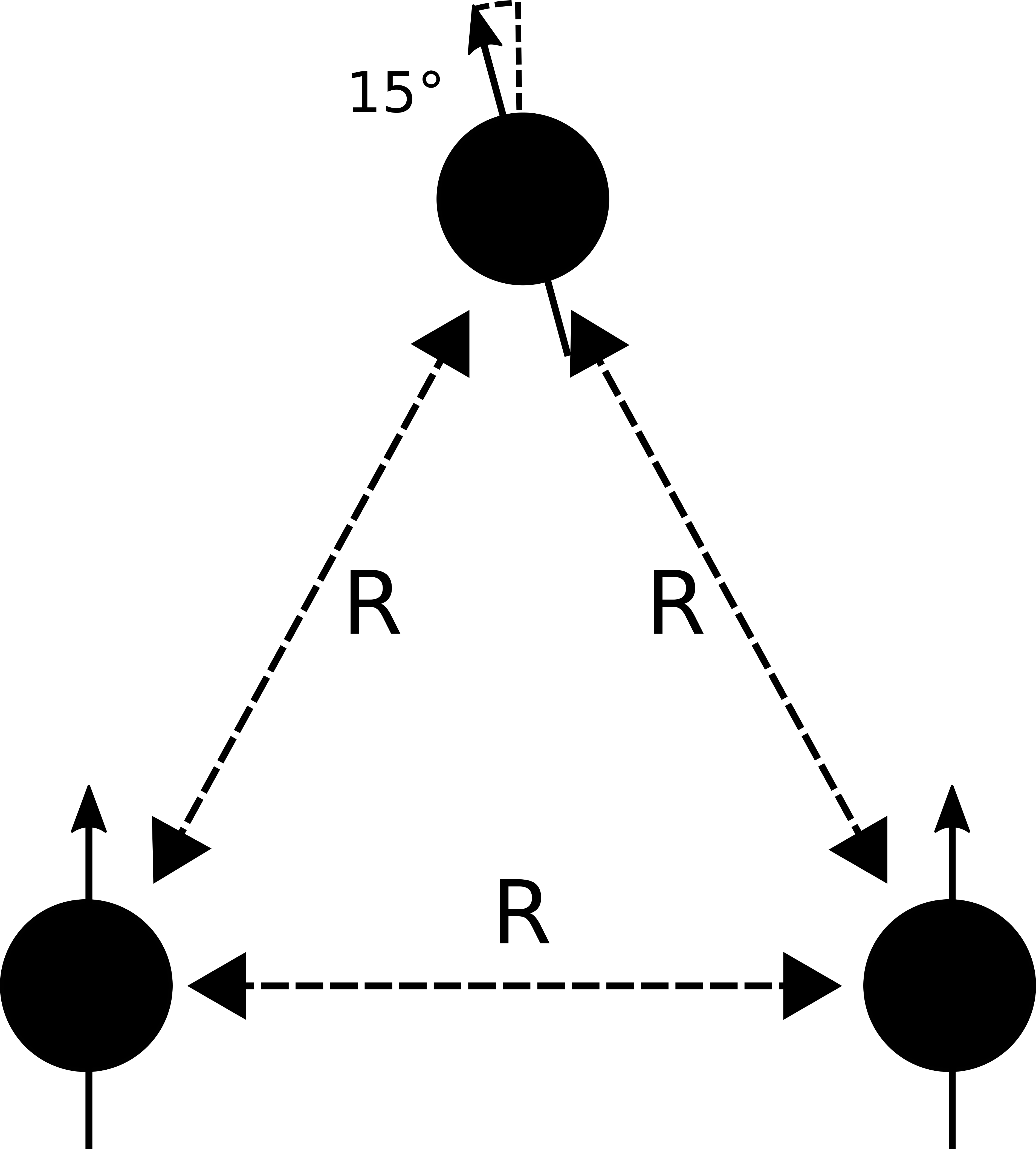}
        \caption{}
        \label{fig:Threeatomtriangularsetup}
    \end{subfigure}%
        \begin{subfigure}[b]{0.33\textwidth}
        \centering
         \includegraphics[height=1.6in]{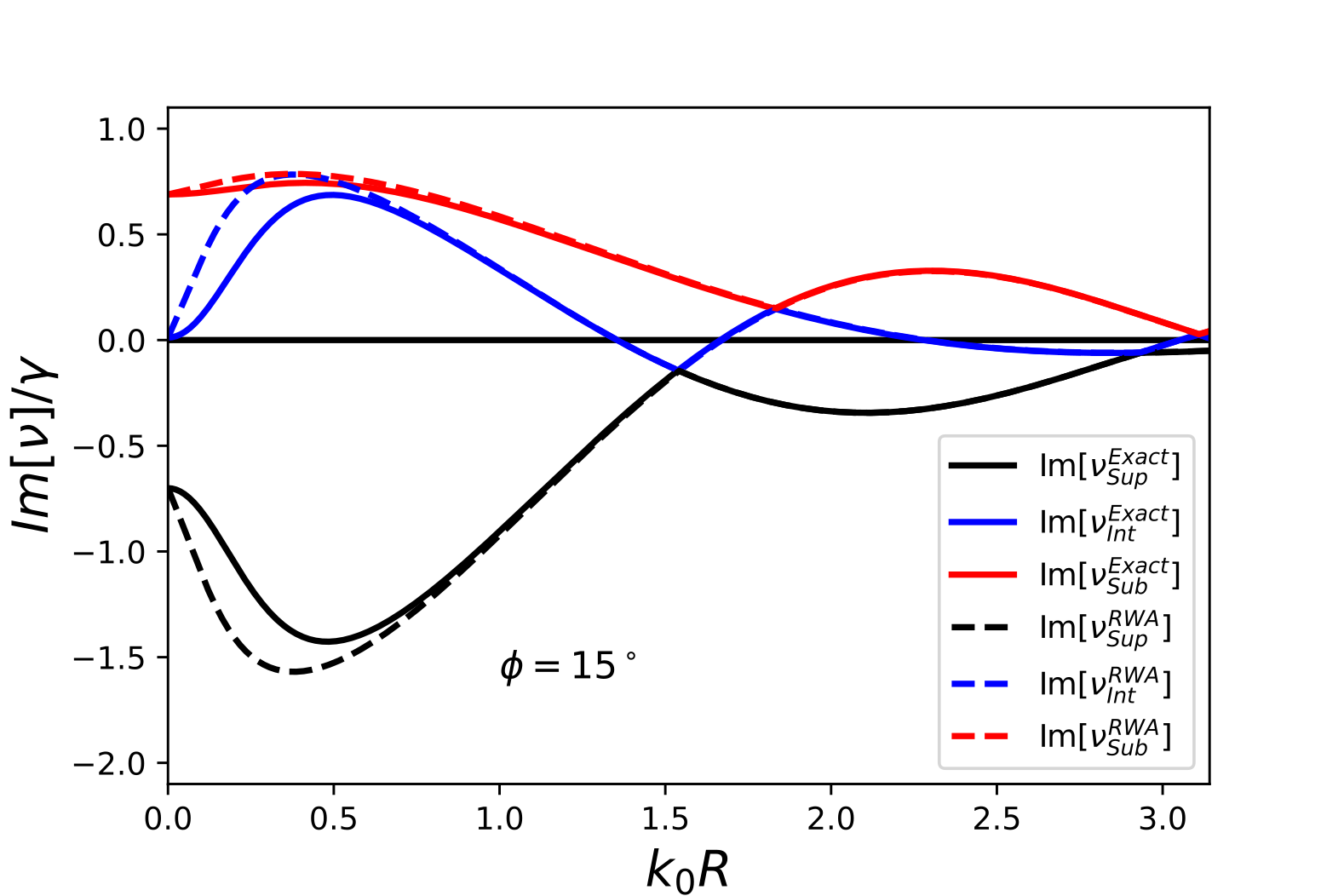}
        \caption{}
        \label{fig:ThreeAtomExactRWAEigs}
    \end{subfigure}
    \begin{subfigure}[b]{0.33\textwidth}
        \centering
         \includegraphics[height=1.6in]{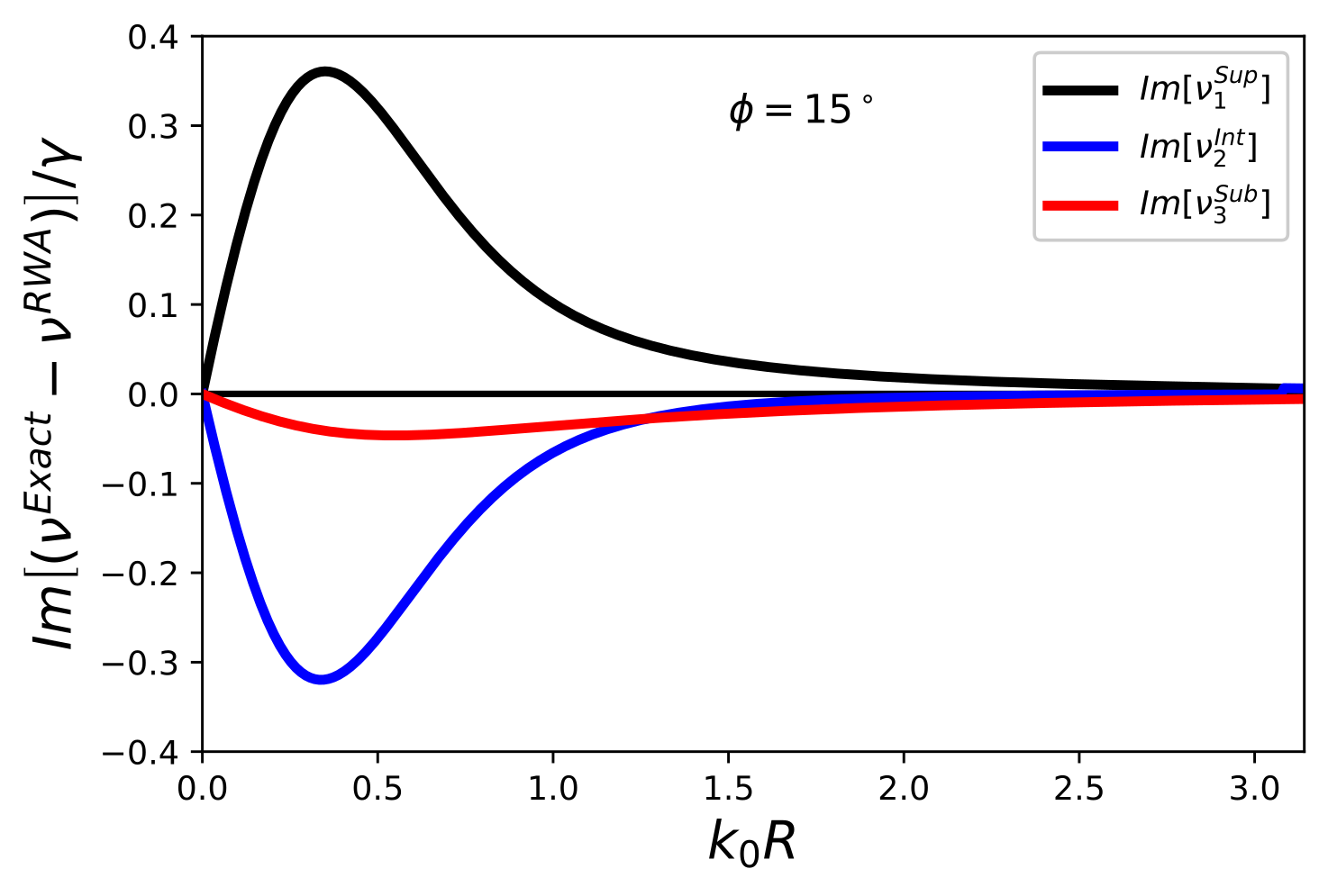}
        \caption{}
        \label{fig:ThreeatomRelativeEigDiffs}
    \end{subfigure}
    \caption{(a) The equilateral triangular setup with in-plane dipole moments, one of which is rotated by $15^\circ$ with respect to the others. (b) Evolution of the decay rates of the three different modes. We see a superradiant mode, an intermediate mode and a subradiant mode. Solid lines correspond to modes  found with the full formalism, while dashed lines are found with the RWA. (c) The relative difference between the modes in the full formalism and RWA. Note that the first axis extends further in (b) than in (c).}
    \label{fig:ThreeAtomSetup}
\end{figure*}

\noindent As a final configuration, we will examine the case of three identical atoms. As we will see, there are qualitatively new effects of making the RWA when going from two to three emitters. 
From Eqs.~(\ref{eq:NAtomSourceTerm}) and (\ref{eq:GeneralizedNAtomPropagator}), it is clear that the collective optical resonances of $N$ atoms is described by the poles of the  $N$-emitter T-matrix, defined in Eqs.~(\ref{eq:NAtomScatteringMatrix})-(\ref{eq:NAtomScatteringMatrixMMatrix}).
This T-matrix can be rewritten as 
\begin{equation}\label{eq:NAtomScatteringMatrixRewritten1}
    \mathbf{T}^{(N)}=\frac{1}{\text{det}\left(\mathbf{M}\right)} \text{diag}\left(T_1, T_2,...,T_N\right)\cdot\text{adj}\left(\textbf{M}\right),
\end{equation}
where $\text{adj}\left(\dots\right)$ is the adjugate matrix and we have used a common rewriting of the inverse of a matrix, $\textbf{M}^{-1}=[{\text{det}\left(\mathbf{M}\right)}]^{-1} \text{adj}\left(\textbf{M}\right)$.
It may seem that Eq.~(\ref{eq:NAtomScatteringMatrixRewritten1})  has resonances for each of the single-atom scattering matrices, and also when $\text{det}\left(\mathbf{M}\right)=0$.
However, for a given $\omega$ coinciding with the resonances of any combination of single-atom scattering matrices $T_n$,
the denominator $\text{det}\left(\mathbf{M}\right)$ contains a term that diverges equally fast or faster.
This is not a general feature of a determinant and its adjugate, but a result of the fact that the single-atom scattering matrices are the same for every row of a given column, such that the determinant contains terms in which each single-atom scattering matrix appears once, while the elements of the adjugate matrix always lack at least one of the single-atom scattering matrices. 
The net effect is that the single-atom resonances do not show up as collective resonances in $\mathbf{T}^{(N)}$. 
So all collective resonances of $\mathbf{T}^{(N)}$ are given instead by the condition $\text{det}\left(\mathbf{M}\right)=0$.  

If we multiply each column of $\mathbf{M}$ with the corresponding single-atom resonance, corresponding to $\omega - \Omega_n - X_n$ in the RWA (while in the full formalism, the equivalent expression is $\omega^2 - \Omega_n^2 -2 \Omega_n X_n$~\cite{Wubs2004a}), we get rid of the single-atom resonances in $\mathbf{M}$.
This operation merely scales the determinant by the same factor for each column that we multiply, and we get
\begin{eqnarray}
   && M'_{mn}=(\omega - \Omega_n - X_n)M_{mn}\\
    &&= (\omega - \Omega_n - X_n)\delta_{mn}- (1-\delta_{mn})J_{mn}\left(\frac{\mu_n}{\mu_m}\right),        \nonumber
\end{eqnarray}
For identical atoms, as we consider here, our problem is reduced to finding the solutions of
\begin{eqnarray}
\text{det}\left[\mathbf{M}'\right]=\text{det}\left[ (\omega - \Omega - X)\mathbb{1} - \mathbf{J}'\right] =0,
\end{eqnarray}
with the interatomic interaction matrix defined as 
\begin{equation}\label{eq:JPrimeMatrix}
    J'_{mn}=(1-\delta_{mn})J_{mn}.
\end{equation}
This is an eigenvalue problem with the resonances given by $\omega_l=\Omega+X+\nu_l$, where $\nu$ is the $l$'th eigenvalue of the matrix $\mathbf{J}'$.

There are three distinct cases to be considered here: all three interactions equal, only two interactions equal, and all interactions different.  Let us first consider collective resonances for the maximally symmetric configuration: in the case of an equilateral triangular formation where also the dipole vectors exhibit the discrete rotational symmetry (all out of plane for example), then all pairwise interatomic interactions are identical, i.e. $J_{12}=J_{13}=J_{23}\equiv J$. The eigenvalues are then solutions of the equation $\nu^3-2J^3-3J^2\nu=(\nu+J)^2(\nu-2J)=0$, and they are therefore linear in J with real coefficients. This means that the decay rate of the $l$'th mode will be unaffected by the RWA, as the real-valued error will not be mixed into the imaginary part of the eigenvalues. These eigenvalues are in agreement with those found in Ref.~\cite{Feng:2013a}.

Next we turn immediately to the  case with the lowest symmetry, of three identical emitters where all three interactions are different. In particular, we consider the symmetry-broken equilateral triangle with in-plane dipole moments, one of which is rotated with respect to the other two, as depicted in Fig.~\ref{fig:Threeatomtriangularsetup}. For this configuration we illustrate in Fig.~\ref{fig:ThreeAtomExactRWAEigs} the impact of the RWA on the collective decay rates, i.e. the imaginary parts of the eigenvalues of the matrix $J'$ as we vary the dimensionless distance parameter $k_0 R$.
Again we see that both in the far field and in the limit $k_0R\rightarrow0$ there is good agreement between the collective rates as obtained with and without the RWA, but in general there are discrepancies in the near field. Fig.~\ref{fig:ThreeatomRelativeEigDiffs} is based on the graphs of Fig.~\ref{fig:ThreeAtomExactRWAEigs} and shows that the maximal relative error in a collective decay rate due to making the RWA exceeds 30\%, and that this maximum occurs in the near field.

In the intermediate case where two out of three interactions are equal, the RWA still causes errors in the collective decay rates. For example, if there was no rotation by $15^\circ$ in Fig.~\ref{fig:Threeatomtriangularsetup}, then we find that the magnitude of the relative RWA errors analogous to Fig.~\ref{fig:ThreeatomRelativeEigDiffs}  are still finite but an order of magnitude smaller (not shown). Mathematically, for this intermediate-symmetry case, the $J'$-matrix is similar to that of  three emitters equally spaced on a line, with all dipole vectors parallel.

After considering these three cases for three-emitter collective decay, we find that in general the RWA affects the collective decay rates of three or more identical emitters. So we find ourselves disagreeing with Ref.~\cite{Feng:2013a} on this point.  Effects of the RWA on the collective decay were also found by Svidzinsky et al.~\cite{Svidzinsky2010a}, for a dense cloud of emitters, complementing our results for few emitters.

\begin{figure}[t!]
\includegraphics[width=\linewidth]{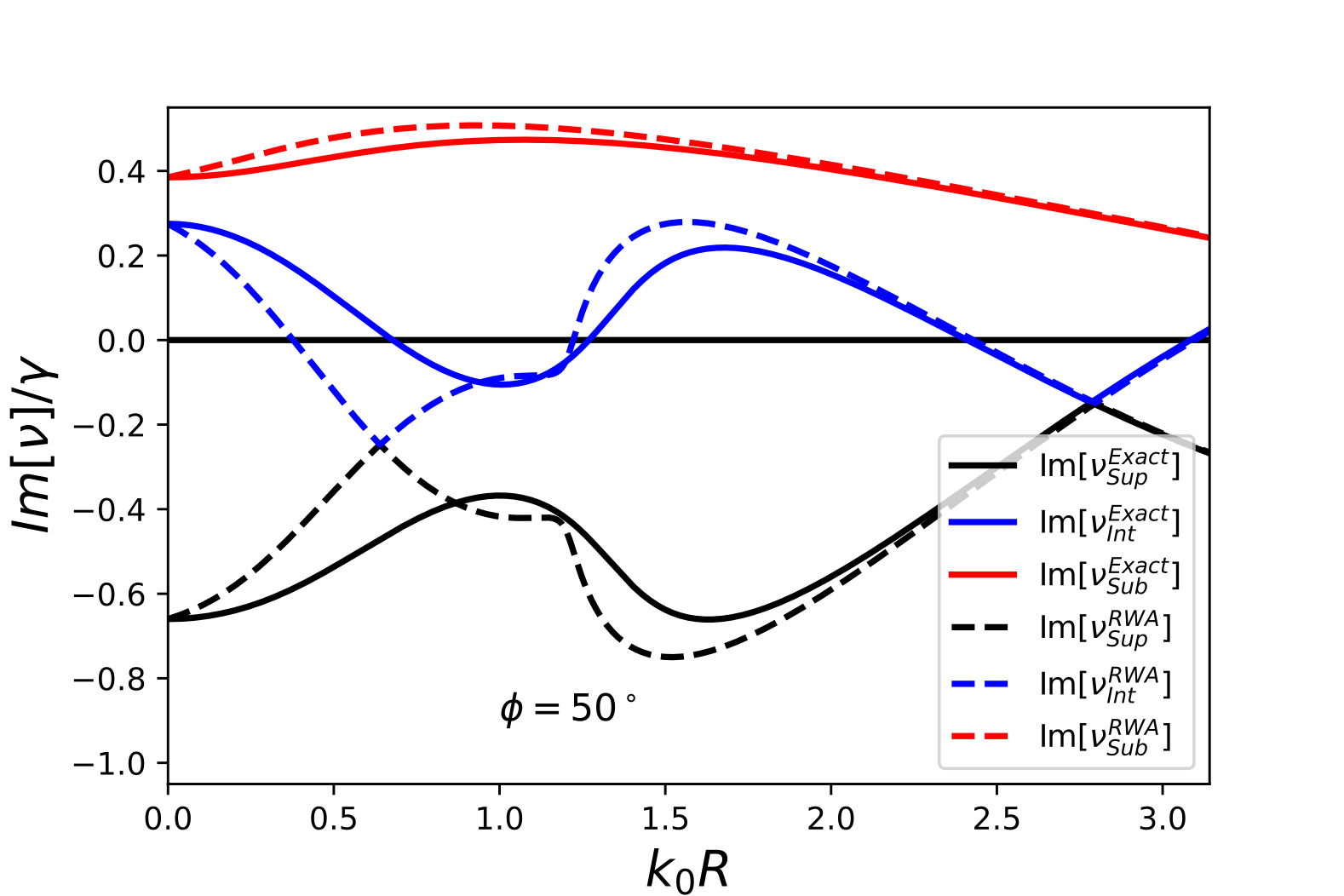}
  \caption{Collective decay rates, scaled by the single-emitter decay rate, for three identical emitters. The setup is as in Fig.~\ref{fig:ThreeAtomSetup} but with the angle of the upper dipole  rotated by $50^\circ$ with respect to the other two instead of $15^\circ$. The decay rates in the full formalism feature an avoided crossing near $k_0R\approx1$, whereas the RWA decay rates have an actual crossing near $k_0R\approx0.7$. This shows that the distance-dependent eigenvalues can be topologically different in the two formalisms.}
  \label{fig:ThreeAtomExactRWAEigs50degrees}
\end{figure}

Another interesting three-emitter phenomenon becomes apparent if we consider the case where the angle of the rotated dipole is $50^\circ$.
In this configuration, the topology of the imaginary parts of the eigenvalues differs between the full formalisms, where we see an anti-crossing near $k_0R\approx1$, and the RWA, where we see a crossing near $k_0R\approx0.7$, as seen in Fig.~\ref{fig:ThreeAtomExactRWAEigs50degrees}.
We will not pursue this result further, but it underlines the shortcomings of the RWA in describing the decay rates of cooperative systems with sub-wavelength separation.

Four emitters can be arranged in an equilateral triangular pyramid, but for configurations with five or more particles, no  constellations exist where all inter-emitter interactions are equivalent, like that of the equilateral triangle with out-of-plane dipole moments. That may suggest that for five or more emitters, the RWA will always affect the collective decay rates, but that is not so: in Sec.~\ref{sec:RingSystems} we will present a class of $N$-emitter systems that  exhibit no RWA errors in any of the $N$ collective decay rates.

\section{RWA for ring systems}\label{sec:RingSystems}
For a  system of $N$ emitters with discrete rotational symmetry, Eq.~(\ref{eq:JPrimeMatrix}) takes the form of a symmetric, cyclic band matrix.
For emitters confined to a plane, these configurations consist of the regular polygons, at least if the dipole vectors have the same ring symmetry. Such regular polygon configurations have already been attracting interest \cite{AsenjoGarcia:2017b,morenocardoner:2020a}.
Due to the discrete rotational symmetry, we can write the interaction between two emitters as an arbitrary linear combination of all possible terms with this rotational symmetry,
\begin{equation}\label{eq:RingDiscreteRotationSymmetry}
    J_{mn}=\braket{n|J|m}=\sum_{\sigma=0}^{N-1} P_\sigma e^{2\pi i (m-n) \sigma/N},
\end{equation}
for some complex coefficients $P_\sigma$.
For convenience we label the first particle with "$0$".
The above property and Eq.~(\ref{eq:JPrimeMatrix}) are given in the "site"-basis where $m$ and $n$ refer to emitter labels. 
We can also introduce the Fourier $k$-basis (i.e. Bloch modes)
\begin{subequations}
\begin{equation}
\ket{k}=\frac{1}{\sqrt{N}}\sum_{n=0}^{N-1} e^{2 \pi i n k/N}\ket{n},
\end{equation}
\text{and the inverse transformation,} 
\begin{equation}
\ket{n}=\frac{1}{\sqrt{N}}\sum_{k=0}^{N-1} e^{-2 \pi i n k/N}\ket{k}.
\end{equation}
\end{subequations}
It is easy to show that the interaction is diagonal in this basis,
\begin{equation}\label{eq:RingLongDerivation1}
\begin{split}
    \widetilde{J}_{kk'}=\braket{k|J|k'} = N P_k \delta_{k,k'},
\end{split}
\end{equation}
with eigenvalues given by the set $\widetilde{J}_{kk}$.
If we insert this result back into Eq.~(\ref{eq:RingDiscreteRotationSymmetry}), we get the 1D Fourier series
\begin{equation}\label{eq:RingSiteEigenValueFourier0}
\begin{split}
     J_{mn}&=\sum_{k=0}^{N-1} \frac{1}{N} \widetilde{J}_{kk} e^{2\pi i (m-n) k/N} \\
     &=\sum_{k=0}^{N-1} \frac{1}{N} \widetilde{J}_{kk} e^{-2\pi i (m-n) k/N}, 
\end{split}
\end{equation}
where we have used the symmetry of the interaction, $J_{mn}=J_{nm}$.
Since each emitter in these ring systems must, by definition, experience every type of interaction, we can fix one index without loss of generality - only the number of sites between two emitters matters.
We fix $m=0$ and only consider the interactions felt by this emitter, as it is representative of all emitters in the ring.
Due to the cyclic nature, we have $J_{0,n}=J_{0,N-n}$. 
The inverse Fourier series to Eq.~(\ref{eq:RingSiteEigenValueFourier0}) using this convention is given by
\begin{equation}\label{eq:RingSiteEigenValueFourier2}
  \widetilde{J}_{kk} = \sum_{n=0}^{N-1} J_{0n} e^{-2 \pi i nk/N},  
\end{equation}
Using a rewriting of the summation in Eq.~(\ref{eq:RingSiteEigenValueFourier2}), and the introduction of a new summation variable $n'=N-n$, it is possible to express the eigenvalues, $\widetilde{J}_{kk}$ as a sum of terms linear in $J_{0n}$ with real coefficients,
\begin{subequations}
\begin{equation}\label{eq:RingFinalEven}
\begin{split}
 \widetilde{J}_{kk}=(-1)^k J_{0\frac{N}{2}} +& 2\sum_{n=1}^{N/2-1} J_{0n} \cos{\left(2 \pi  nk/N\right)},
 \end{split}
\end{equation}
for N even, and
\begin{equation}\label{eq:RingFinalOdd}
 \widetilde{J}_{kk}=2 \sum_{n=1}^{(N-1)/2} J_{0n} \cos{\left(2 \pi  nk/N\right)}, 
\end{equation}
for N odd.
\end{subequations}
We have made use of the fact that the diagonal of interaction matrix in the site-basis is zero, see Eq.~(\ref{eq:JPrimeMatrix}).
This linear relation between the eigenvalues and the interatomic interactions shows that for $N$ identical emitters in a ring configuration, none of their $N$ collective decay rates, proportional to $\mbox{Im}[\widetilde{J}_{kk}]$, are affected by changes in the real parts of the interactions. So, for this narrow class of configurations, there will be no error in the collective decay rates associated with making the RWA.

\section{Summary and conclusion}\label{Sec:summary}

\noindent We have utilized a multiple-scattering formalism in order to compare quantitatively the emission properties of single-,  two- and many-atom emission in formalisms with and without the RWA approximation in the light-matter interaction. We considered arbitrary inhomogeneous dielectric media and  approximated the quantum emitters as quantum harmonic oscillators. 
Whenever there is more than a single emitter, the validity of the RWA cannot be read off directly from the  Hamiltonian Eq.~(\ref{eq:TotalSymbolicHamiltonian}), which features emitter-field interactions but no direct emitter-emitter interactions, as also mentioned in~\cite{Leonardi1986a}. Here we show how in our multiple-scattering formalism the validity of the RWA can be tested by comparing instead the induced inter-emitter interactions (proportional to a field propagator) with and without making the RWA. These induced interactions emerge in the implicit solution of the dynamics as described by the Lippmann-Schwinger equation~(\ref{eq:RWAFieldOperatorPlusMinusFreeSourceScattering}).

Without making the RWA in our quantum optical model, the propagator that emerges is the classical electromagnetic Green function~\cite{Wubs2004a}, while here we find that within the RWA the propagator that emerges carries an error term. It is a subtle error in the sense that it is purely real-valued for any non-lossy photonic medium. Therefore the RWA error leaves the single-atom decay rate unchanged. And while  the induced two-atom interaction $J_{12}$ is affected by making the RWA,  the super- and subradiant decay rates are still unaltered, irrespective of their distance. By contrast, the predicted two-atom sub- and superradiant level shifts are affected by making the RWA.

In case of two detuned quantum emitters, the real-valued  error of the RWA propagator does  affect both collective decay rates, and the errors 
can become significant. In Fig.~\ref{fig:2AtomDetunedResonance} we found that the largest relative RWA errors in the collective decay rates occur at intermediate distances,  in between the $R \ll \lambda$ (Dicke) limit and the far-field limit $R \gg \lambda$. Unlike for the lossless media considered here, we expect that for lossy media  collective emission rates will be affected by making the RWA, even for identical emitters.

For three or more quantum emitters, we found that the situation is qualitatively different than for a pair: for three identical emitters, the collective decay rates are in general affected by making the RWA, while for two identical emitters this was not the case. Our study of few-emitter emission provided insight, and complements the numerical study of Ref.~\cite{Svidzinsky2010a} where effects of the RWA were studied for dense clouds of atoms.   

The fact that the RWA produces the same single-atom spontaneous-emission rates has sometimes been explained by the emission being an on-shell process. Collective single-photon decay by two identical atoms could again be called an on-shell process, and indeed the RWA does not change the rates. But then comes single-photon decay by three or more identical atoms, again an on-shell process, where the collective decay rates are affected by the RWA. This shows that the on-shell or off-shell nature of the transition is not the decisive distinction for the accuracy of the   collective decay rates within the RWA. 

While in general the collective decay rates are affected by the RWA, there exist special configurations of identical emitters for which this is not the case, even under a common scaling transformation of all distances involved. Mathematically, the requirement is that changes in the real parts of the interatomic interactions do not affect the imaginary parts of the eigenvalues of the complex interaction matrix $J'$ of Eq.~(\ref{eq:JPrimeMatrix}). We identified cyclic symmetric matrices for which this is indeed the case: a Fourier series analysis showed that the eigenvalues are given by linear combinations of the complex-valued interactions, with (real-valued) cosines as coefficients. 
We expect but have not proven that this is the only type of scalable multi-emitter configuration where the RWA does not affect the collective decay rates. In general, the imaginary parts of the eigenvalues depend in a nonlinear fashion on the elements of the interatomic interaction matrix, and the common configuration of identical quantum emitters positioned equidistantly on a line is an important example where the RWA does change the collective decay rates. For non-identical emitters, even circular configurations will not prevent errors in the decay rates by making the RWA.  

The relative error on the real part of the RWA propagator was shown to diverge for the scalar propagator in the limit $k_0R\rightarrow0$, while tending towards a factor of $1/2$ for the dyadic propagator. The accuracy of the RWA for light thus strongly depends on light being described by  vector rather than scalar waves. This RWA error of fifty percent is already almost reached at finite distances typical for F{\" o}rster resonance energy transfer. The factor $1/2$ implies that at very short distances, the non-rotating terms become as important as the rotating ones. This value is supported by Friedberg and Manassah, who also  found the equivalent contribution of resonant and off-resonant terms in the near-field~\cite{FRIEDBERG:1973101}. We have not found a simple explanation for the factor 1/2 that at the same time explains why for scalar waves the relative RWA error diverges. 

Our calculations illustrate that interatomic interactions at shorter distances are mediated by a wider interval of modes, below some distance also requiring the non-rotating terms. This can be understood from a energy-time-uncertainty argument and for free space by a  Fourier relation between distance and wave vector. The usual argument for neglecting the anti-rotating terms fails for emitters at sub-wavelength distances: after going to a rotating frame, the rapidly varying time-dependent factors of the anti-rotating terms do {\em not} average to zero on all relevant time scales of the physical problem, because the tiny travel time $R/c$ between the two nearest  emitters is one such relevant time scale. 
Our results clarify and quantify how much the RWA affects collective light emission, and when Green-function based formalisms are to be preferred.

\begin{acknowledgments}
\noindent The authors would like to thank Igor Protsenko,  Emil Vosmar Denning, Kristian Seegert, Devashish Pandey, Nicolas Stenger, and Sanshui Xiao  for stimulating discussions. M.W. acknowledges support by the Danish National Research Foundation through NanoPhoton - Center for
Nanophotonics (grant number DNRF147) and Center
for Nanostructured Graphene (grant number DNRF103), and from the Independent Research
Fund Denmark - Natural Sciences (project no. 0135-004038).
\end{acknowledgments}


\bibliography{Mads_RWA}%
\end{document}